%% file: main.tex
\documentclass[letterpaper, 11pt]{extarticle}
\usepackage[english]{babel}
\usepackage[margin = 1in]{geometry}
\usepackage{amsmath, amssymb}
\usepackage{algorithm}
\usepackage{algpseudocode}
\usepackage{multirow}
\usepackage{makecell}
\usepackage{mathrsfs}
\usepackage{amsfonts}
\usepackage{amsmath}
\usepackage{amsthm}
\usepackage{amssymb}
\usepackage{physics}
\usepackage{dsfont}
\usepackage{esint}
\usepackage{enumerate}
\usepackage[shortlabels]{enumitem}
\usepackage{framed}
\usepackage{csquotes}

\usepackage{float}
\usepackage{tabularx}
\usepackage{xcolor}
\usepackage{multicol}
\usepackage{subcaption}
\usepackage{caption}
\usepackage{enumitem}
\setlist[itemize]{itemsep=0.05em, topsep=0.3em, parsep=0em}
\setlist[enumerate]{itemsep=0.05em, topsep=0.3em, parsep=0em}
\usepackage[round, authoryear]{natbib}
\usepackage{graphicx}
\usepackage{bm} 

\usepackage{hyperref}
\definecolor{links}{rgb}{0.36,0.54,0.66}
\hypersetup{
	colorlinks = true,
	linkcolor = black,
	urlcolor = blue,
	citecolor = blue,
	filecolor = blue,
	pdfauthor = {Author},
	pdftitle = {Title},
	pdfsubject = {subject},
	pdfkeywords = {one, two},
	pdfproducer = {LaTeX},
	pdfcreator = {pdfLaTeX},
}

\usepackage{titlesec}
\usepackage[many]{tcolorbox}

\titlespacing*{\chapter}{0cm}{-2.0cm}{0.50cm}
\titlespacing*{\section}{0cm}{0.50cm}{0.25cm}

\newtcbtheorem[]{problem}{Problem}%
{enhanced,
	colback = black!5, %white,
	colbacktitle = black!5,
	coltitle = black,
	boxrule = 0pt,
	frame hidden,
	borderline west = {0.5mm}{0.0mm}{black},
	fonttitle = \bfseries\sffamily,
	breakable,
	before skip = 3ex,
	after skip = 3ex
}{problem}

\tcbuselibrary{skins, breakable}

\makeatletter
\newcommand*\bigcdot{\mathpalette\bigcdot@{.5}}
\newcommand*\bigcdot@[2]{\mathbin{\vcenter{\hbox{\scalebox{#2}{$\m@th#1\bullet$}}}}}
\makeatother

\usepackage{float}
\usepackage{tikz}
\usetikzlibrary{shapes.geometric, arrows.meta, positioning, calc}
\usepackage{authblk}
\begin{document}

\title{\textbf{BAYESIAN ASSURANCE FOR SAMPLE SIZE DETERMINATION IN ACCELERATED RELIABILITY DEMONSTRATION TESTING WITH LOCATION EFFECTS}

}
	% Authors and affiliation
	\author[1]{Juliana Ruth Okodoa\thanks{Corresponding author. Email: j.r.okodoa2@newcastle.ac.uk (Juliana Ruth Okodoa)}}
	\author[1]{Kiran Prajapat}
    \author[1]{Aleksandra Svalova}
    \author[1]{Kevin J. Wilson}

	\affil[1]{\small School of Mathematics, Statistics and Physics, Newcastle University, Newcastle upon Tyne, United Kingdom, NE1 7RU.}
	\date{}
	\maketitle
	
	\begin{abstract}
        \noindent Reliability demonstration tests are designed to provide statistical evidence that a product meets a pre-specified reliability requirement before market release. Traditional approaches rely on frequentist criteria and offer no direct measure of the probability that a planned test will lead to a successful outcome at the design stage. Bayesian assurance addresses this by quantifying, before any data are collected, the probability that the test will demonstrate the required reliability.
        This study develops a sample size determination procedure for constant-stress accelerated life testing. Unit lifetimes follow a Weibull distribution with a log-linear stress-life relationship, and random production location effects are incorporated into this relationship to account for manufacturing heterogeneity. A prior elicitation procedure is developed in which expert assessments of observable lifetime quantities are systematically converted into prior distributions for all model parameters.
        The methodology is illustrated using data from the Personal Alert Safety System (PASS) of a self-contained breathing apparatus (SCBA), tested by a commercial manufacturer of industrial safety equipment. The results show that the required sample size is strongly dependent on both the reliability requirement and the expert judgments used to construct the prior distribution. Planning from historical data alone, without accounting for parameter uncertainty, underestimates the number of units needed and increases the risk that the test fails to demonstrate the requirement.  
    \end{abstract}
		
    \noindent\textbf{Keywords:} Hardware product development; prior elicitation; reliability analysis; Bayesian; Weibull model; Eyring model; random effects; product validation and verification. \\[3pt]    
    \noindent \textbf{MSC 2020:} 62N05, 62F15, 62K05, 62P30, 62N01      

    \section{Introduction} 
    \label{sec:intro}
    
    Safety-critical products must demonstrate that they satisfy specified reliability requirements before they can be released for use. An important example is the self-contained breathing apparatus (SCBA), which provides respiratory protection for firefighters and other personnel operating in hazardous environments. The reliability of such equipment is critical, since product failure can have fatal consequences for user safety. Manufacturers therefore require appropriate reliability demonstration testing (RDT) to provide statistical evidence that the product meets its specified reliability target. In an RDT, a predetermined number of units are tested, and a pre-specified acceptance rule, based on the observed failures or failure times, determines whether the reliability requirement has been demonstrated (\citet{elsayed2012book, hamada2008, meeker2022}). For highly reliable products, however, failures may be rare under normal operating conditions, making it difficult to obtain sufficient lifetime information within a practical testing period. Accelerated life testing (ALT) addresses this problem by exposing units to elevated stress levels and using a physically motivated stress--life relationship, such as the Eyring model, to extrapolate reliability to the intended use condition (\citet{Meeker1998, nelson1990}). This creates an important planning question: given the accelerated testing conditions and the available historical information, how many units should be tested to provide a sufficiently high probability of demonstrating the required reliability?
    
    Traditional approaches to RDT sample-size determination are generally based on the power of a statistical test or on criteria that control the producer's and consumer's risks of reaching an incorrect conclusion (\citet{hamada2008, wilsonandfarrow2021}). Although these approaches provide established procedures for evaluating test plans, they typically require the specification of reliability levels at which the product is considered acceptable or unacceptable. Such values may be difficult to establish in practice, particularly when uncertainty about the product's underlying reliability is substantial. Bayesian reliability assurance tests provide an alternative risk-based framework. For example, \citet{Hamada2014} developed binomial assurance procedures for system reliability, demonstrating how information from component testing can reduce the number of system-level tests required to satisfy specified risk criteria. \citet{Weaver2017} extended this approach to degradation data by incorporating prior information from related degradation studies. Further developments have considered reliability demonstration in systems with multiple levels of information (e.g.number of units to test, consumer's risk, producer's risk, etc.) and the influence of additional sources of variability on test planning (\citet{Fernandez2023, Li2017}). Cost considerations have also been incorporated into RDT planning, including accelerated degradation testing and multi-objective test-plan optimisation (\citet{kimwilson2025, lu2016}). Nevertheless, risk-based approaches do not necessarily provide a direct criterion for selecting a sample size according to the overall probability of successfully demonstrating a reliability requirement.
    
    Assurance provides such a criterion by quantifying the probability that a planned test will lead to a successful decision before any new data are collected. The concept originated in Bayesian approaches to sample-size determination in clinical trials, where the probability of a successful trial is evaluated by averaging over uncertainty in the underlying parameters (\citet{ohagan2005, Spiegelhalter1986}). Unlike conventional power, which is calculated conditional on a single assumed parameter value, assurance integrates the probability of success over a prior distribution, thereby incorporating parameter uncertainty directly into the design (\citet{ohagan2005}). Simulation-based Bayesian design also distinguishes between the distribution used to generate plausible data at the planning stage and the statistical model used to analyse the resulting observations (\citet{wanggelfand2002}). In reliability engineering, \citet{wilsonandfarrow2021} developed assurance-based RDT procedures for binomial and Weibull likelihoods. Their framework determines the sample size required to achieve a specified probability of successfully demonstrating a reliability requirement, with the design prior representing the manufacturer's beliefs or available historical information and the analysis prior specifying the basis for the subsequent assessment. 
    
    An important extension is required when reliability evidence is obtained under accelerated conditions and subsequently extrapolated to the intended use condition. In this setting, the probability of a successful demonstration depends not only on the lifetime distribution but also on uncertainty in the stress--life relationship used for extrapolation. Moreover, industrial testing programs may involve units originating from different manufacturing sites or production locations, where systematic differences in production and assembly processes can lead to variation in lifetime behavior. Such variation is not necessarily explained by the applied stress alone. Previous reliability studies have considered random group, block and test-chamber effects to account for dependence and heterogeneity between experimental units (\citet{freeman2010, kensler2014, leon2007, leon2009, seopan2018}). These studies demonstrate the relevance of accounting for grouping structures when modeling reliability data and predicting product performance. In the present work, location heterogeneity is incorporated through a random location effect in the Weibull--Eyring model. This allows differences between test locations to contribute to the uncertainty in reliability predictions at the use condition and, consequently, to the evaluation of assurance. The focus is therefore on extending assurance-based RDT planning to a setting in which both accelerated lifetime behaviour and location-related variation influence the demonstration decision.
    
    The use of historical information introduces a further consideration in assurance-based test planning: how should previously observed data inform the design of a future RDT? Two approaches are considered in this study. The first, which we propose, uses a design posterior, obtained by updating an elicited prior distribution with the likelihood from historical accelerated life test data. This approach incorporates the historical observations while retaining uncertainty about the underlying model parameters. The second is an empirical approach based on resampling the observed historical units, using the empirical distribution of the available data to represent the population for future testing. These approaches provide different representations of historical information and may consequently lead to different sample-size recommendations. As we show, the choice matters: ignoring parameter uncertainty leads to a test that is too small.
    
    The specification of the initial prior is particularly important when historical information is incorporated through Bayesian updating. Expert elicitation offers a systematic means of translating engineering knowledge and experience into probability distributions. Established methods recommend eliciting judgements about observable quantities rather than abstract model parameters, since quantities such as expected lifetimes or reliability at specified stress conditions are generally more interpretable to subject-matter experts (\citet{Bedrick1996, Garthwaite2005, Kadane1998, OHagan2006}). Structured elicitation protocols, including the Sheffield Elicitation Framework, provide further guidance for obtaining and evaluating expert judgements (\citet{Gosling2018}). In the present study, prior information is elicited independently from two experts using a structured procedure based on observable quantities. The resulting prior distributions are used to construct design posteriors, allowing the influence of expert judgements on assurance and the required sample size to be investigated.

    The paper makes four contributions to the field of RDT through developing an assurance-based framework for constant-stress accelerated reliability demonstration testing under a Weibull--Eyring lifetime model with a random location effect, and makes four contributions.  
    First, we extend assurance-based RDT planning to constant-stress ALT by incorporating a random location effect into the log-linear Eyring relationship. The resulting model accounts for location-to-location heterogeneity when extrapolating reliability to the intended use condition and evaluating the probability of a successful demonstration. Second, we elicit priors from two experts through a structured procedure based on observable quantities of the accelerated lifetime model, and use them to examine the sensitivity of the required sample size to differences in expert judgement. Third, the proposed and empirical approaches to incorporating historical ALT data into the design are compared, showing that the empirical approach, which ignores parameter uncertainty, underestimates the required sample size. 
    Finally, the proposed methodology is applied to the SCBA PASS device reliability demonstration testing, illustrating how accelerated lifetime evidence, location heterogeneity, and expert-informed prior information can be brought together to determine an appropriate sample size for demonstrating reliability.
    
    The remainder of the paper is organized as follows. Section~\ref{sec:industrial_background} describes the SCBA application and the accelerated life testing data. Section~\ref{sec:model} presents the Weibull--Eyring lifetime model incorporating the random location effect, with Section~\ref{sec:prior} detailing the prior elicitation procedure. Section \ref{sec:bayesian_assurance} develops the Bayesian assurance methodology and the alternative approaches to incorporating historical information. Section \ref{sec:simulations} presents the simulation study and associated results. Finally, Section \ref{sec_7} provides the conclusions and discusses the implications of the proposed framework for reliability demonstration testing.

    \section{Industrial Application Background }%/ Data Description} 
    \label{sec:industrial_background} 
	The data were obtained from an industrial safety equipment manufacturer. The product under study is a Self-Contained Breathing Apparatus (SCBA) compliant with EN 137, used in firefighting and emergency rescue operations to provide breathable air in oxygen-deficient, toxic, or immediately dangerous to life or health (IDLH) environments.
The SCBA is designed to operate under demanding environmental conditions, typically between $-$$30^\circ\mathrm{C}$ and $+60^\circ\mathrm{C}$, with additional short-duration exposure to higher radiant heat flux during firefighting activities. In practice, the system is expected to maintain safe and stable performance under rapidly changing thermal conditions, mechanical shock, and high humidity. Given the safety-critical nature of the application, reliability is dominated by the ability of the system to maintain airtight integrity and stable air delivery under thermal stress.

Failures considered in this study are associated with thermal stress–induced degradation mechanisms. 
Three failure modes were identified as the most safety-critical functional degradations that directly compromise breathable air delivery or user protection under high-temperature exposure:
\begin{itemize}
    \item free-flow failures of the lung demand valve, caused by thermal expansion of internal components and relaxation of spring-loaded assemblies at elevated temperatures;
    \item loss of sealing integrity, leading to leakage at interfaces such as O-rings and facepiece seals;  
    \item loss of tensile strength or mechanical integrity of the harness system, due to polymer softening and thermo-mechanical fatigue.
\end{itemize}
 \
Temperature was used as the accelerating stress variable to replicate and intensify thermal conditions experienced in firefighting environments.

In the accelerated life testing of the SCBA, three elevated temperature levels were used; $60^\circ\mathrm{C}$,
$72^\circ\mathrm{C}$, $84^\circ\mathrm{C}$.
These levels were selected to represent increasing severity of thermal exposure, ranging from upper-end operational conditions to highly accelerated stress conditions used to induce failures within practical test durations. A total of $90$ units~ from 2 different production locations were tested, with an equal allocation across stress levels. Each unit was tested under a Type-I right-censored life test with a censoring time of $200$ hours (hrs). 
Units that did not fail within this period were recorded as censored observations. 
The censoring time of $200$ hrs corresponds to $30$ minutes of daily usage over a $10$-year operational life. 

\begin{figure}[!t]
    \centering    \includegraphics[width=\textwidth]{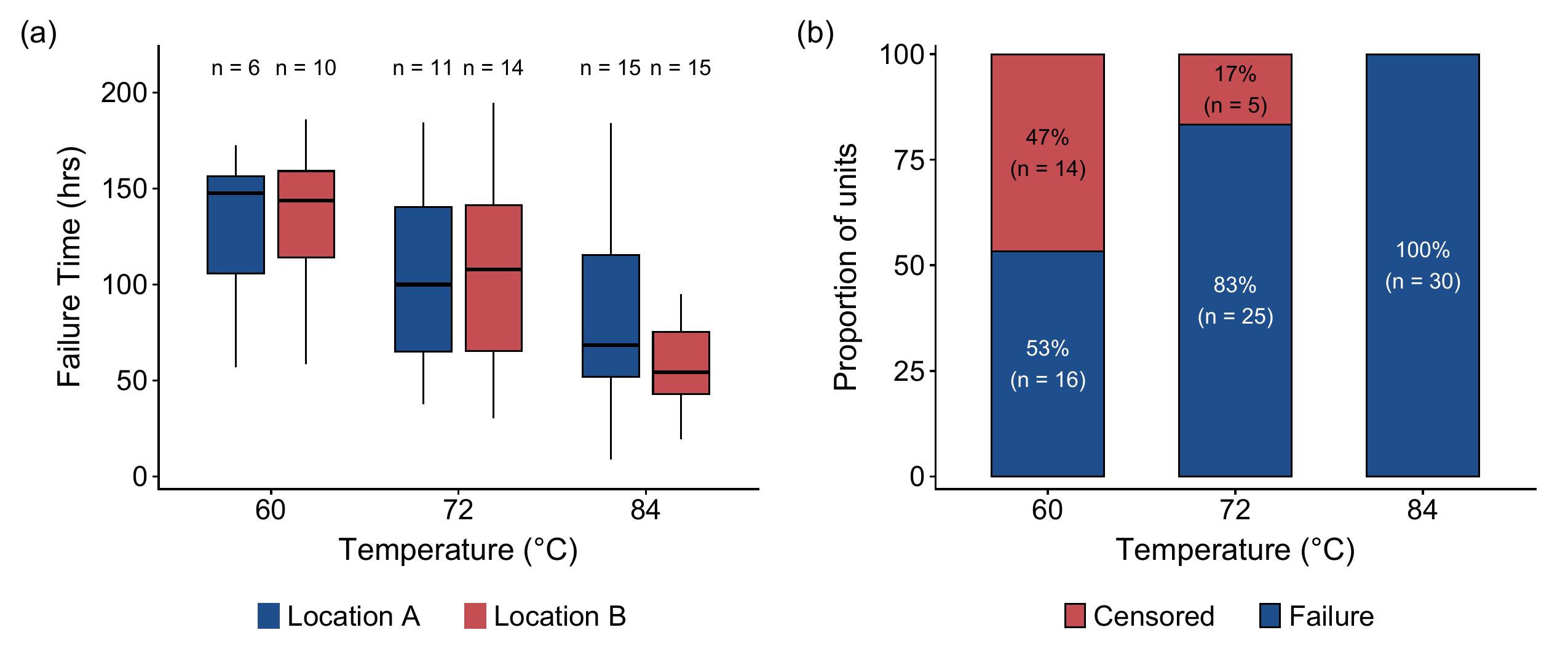}
    \caption{(a) Observed failure times by temperature and location. (b) Proportion of failed and censored units by temperature.}
    \label{fig:combined-plots}
\end{figure}

Figure \ref{fig:combined-plots}(a) presents boxplots of the observed failure times by production location and temperature stress level, and the Figure \ref{fig:combined-plots}(b) to the right shows the proportion of failures and censored observations at each stress level. 

These data are used in the simulation study of Section \ref{sec:simulations} in two ways. In the proposed approach, they are used to obtain the design posterior, from which future test data are simulated. In the empirical approach, used for comparison, subsamples of increasing size are drawn to evaluate the Bayesian assurance at each candidate sample size.

	\section{Model Specification}
	\label{sec:model}
	
	This section presents the statistical model for the proposed ALT-based RDT. The model formulation consists of two components: first, specification of the life-testing experiment under study, model assumptions, and likelihood function (Section \ref{subsec:setup}); and second, a structured prior elicitation procedure that translates the engineer's knowledge into prior distributions for all model parameters (Section \ref{sec:prior}). 

    \subsection{Test Conditions and Model Specification}
    \label{subsec:setup}
    
    Let $n$ units be subjected to a constant-stress accelerated life test, with temperature serving as the accelerating stress factor. Depending on the application, other stress factors may be considered, such as voltage in electrical degradation studies or pressure in mechanical fatigue testing. Each unit $i$ is randomly assigned a stress level $s_i$ ranging between normal use stress $s_0$ and above by the maximum allowable stress $s_h$. Each unit is observed until failure or until
    a pre-specified censoring time $t_C$ is reached, at which point the experiment is terminated. Observations are subject to a common Type-I censoring time $t_C$. 
    
    The test design is specified by:
    \begin{itemize}
        \item $n$ units each assigned a test stress $s_i \in (s_0, s_h]$, $i = 1, \ldots, n$,
        \item A common Type-I right censoring time $t_C$ (here, $t_C = 200$ hrs),
        \item A target use stress $s_0 \leq s^* \leq s_h$ 
        at which the reliable life is required to be assessed.
    \end{itemize}
    
    The censoring status of unit $i$ is recorded using the indicator variable
    \[
    z_i =
    \begin{cases}
    1, & \text{if unit } i \text{ fails by time } t_C,    \\
    0, & \text{if unit } i \text{ is right-censored at }
         t_C,
    \end{cases}
    \]
    and the observed data of the $i$-th unit are the pair $(t_i, z_i)$, where $t_i \leq t_C$ is the failure time. The complete dataset is
    \[
    \mathcal{D}
    = \bigl\{(t_i,\, z_i) : i = 1, 2, \ldots, n
      \bigr\}.
    \]
     In accelerated life testing, the raw stress variable denoted by $s_i^{\text{raw}}$ and used in the statistical model is the original physical measurement but a monotonic  transformation of it. Following \citet{nelson1990}, let $\xi_i$ denote the physical stress characteristic applied to the $i$-th unit, such as temperature in thermal experiments, voltage in electrical degradation studies, or pressure in mechanical fatigue tests, and define the raw stress variable as $s_i^{\text{raw}} = g(\xi_i)$ for a monotonic function $g$ chosen to reflect the underlying degradation mechanism. For thermally accelerated experiments, the standard choice motivated by the Eyring and Arrhenius relationships (\citet{nelson1990}) is $g(T) = 1/T$, where $T$ denotes absolute temperature in Kelvin.

    The raw stress variable $s_i^{\text{raw}}$ is then rescaled as 
    \begin{equation}
        s_i = \frac{s_i^{\text{raw}}}{\bar{s}},
        \qquad
        \bar{s} = \frac{1}{|\mathcal{U}|}
        \sum_{u \in \mathcal{U}}u,
        % \bar{s} = \frac{1}{K}\sum_{k=1}^{K}
        % s_k^{\text{raw}},
        \label{eq:scaling}
    \end{equation}

     where $\mathcal{U}$ denotes the set of distinct values of $s_i^{\text{raw}}$, and $|\mathcal{U}|$ denotes the number of distinct values in that set and $\bar{s}$ is their unweighted mean.  Under this rescaling, $s_i = 1$ corresponding to the mean test stress, similar to \citet{nelson1990, Meeker1998}. 
    
    We assume that random variable $T$, denoting the failure time of an arbitrary unit, follows a Weibull distribution
    with shape parameter $\beta > 0$ and rate parameter
    $\rho(s) > 0$, which depends on stress s; abbreviated as $T \sim \text{Weibull}(\rho(s), \beta)$. Its probability density function (PDF) is given by
    \begin{equation}
        f(t \mid \rho(s), \beta)
        = \rho(s) \beta \left (\rho(s) t \right)^{\beta - 1}
          \exp \left[-\left( \rho(s) t \right)^{\beta}\right],
        \qquad t > 0,
        \label{eq:weibullpdf}
    \end{equation}
    and the corresponding survival function is
    \begin{equation}
        S(t \mid \rho(s), \beta)
        = \exp \left[-(\rho(s) t)^{\beta}\right],
        \label{eq:survivalfct}
    \end{equation} 
    where $(S(t\mid\rho(s),\beta))$ denotes the probability of survival beyond time ($t$), conditional on the stress-dependent rate parameter $(\rho(s))$ and the Weibull shape parameter ($\beta$).
    It is assumed that the lifetime distribution is affected by the applied stress entirely through the rate parameter $\rho(s)$, while the shape parameter $\beta$ remains constant across all stress levels. This is a standard assumption in constant-stress ALT (\citet{nelson1990,Meeker1998}) and ensures that the acceleration structure is determined solely by the stress-life relationship presented below.
    
    The lifetime of unit $i$, tested at stress level $s_i$ and belonging to production group with index $l_i$, is then assumed to follow
    \begin{equation}
        T_i \mid \rho_i, \beta
        \sim \text{Weibull}(\rho_i, \beta),
        \label{eq:weibull}
    \end{equation}
    where $\rho_i = \rho(s_i, {l_i}) > 0$ is the unit-specific rate parameter determined by both the applied stress and the production group of unit $i$, as specified
    below.
    
    For homogeneous failure data from ALT, to model the dependence of the rate parameter on the applied stress, a log-linear stress-life relationship is assumed,
    \begin{equation}
        {\log\left(\rho(s)\right)} = \alpha_0 + \alpha_1 s,
        \label{eq:loglinear}
    \end{equation}
    where $\alpha_0 \in \mathbb{R}$ and $\alpha_1 > 0 $ are the intercept and slope parameters of this log-linear relationship. $\alpha_1 > 0 $ indicates that the failure rate increases monotonically with stress. For the temperature as the
     physical accelerating variable, this corresponds to the Eyring model.
    
    To account for the systematic heterogeneity in baseline failure times arising from variation across locations; manufacturing production sites, batches, or assembly conditions, a random location effect $\epsilon_l$ is incorporated for units belonging to the production group $l = 1, \ldots, L$. Therefore, the complete stress-life log-linear model for the $i$-th unit in a production group $l_i$ becomes 
    \begin{equation}
        \log(\rho_i)
        = \alpha_0 + \alpha_1 s_i + \epsilon_{l_i},
        \label{eq:fullmodel}
    \end{equation}
    so that $\rho_i = \exp \left(\alpha_0 + \alpha_1 s_i  + \epsilon_{l_i}\right)$. Observe that $l_i$ represents the location index of the $i$-th unit under the life test such that $l_i \in \{1,2,\ldots L\}$, and it is assumed that the locations of the units are known in advance.

    In the Bayesian framework adopted here, the location effects $\epsilon_l$ are treated as unknown parameters.
    This is equivalent to the random effects formulation of \citet{leon2007}, but handled using a hierarchical prior rather than marginalization (\citet{seo2017}). The
    complete parameter vector of interest becomes
    \begin{equation}
        \boldsymbol{\theta}
        = \bigl(\boldsymbol{\alpha}^\top, \beta,
          \boldsymbol{\epsilon}^\top\bigr)^\top
        \in \Theta,
        \label{eq:paramvec}
    \end{equation}
    where $\boldsymbol{\alpha} = (\alpha_0, \alpha_1)^\top$, $\boldsymbol{\epsilon} = (\epsilon_1, \ldots, \epsilon_L)^\top$, and the parameter space is
    \begin{equation}
    \Theta = \bigl\{
    (\alpha_0, \alpha_1, \beta,
    \boldsymbol{\epsilon}) :
    \alpha_0 \in \mathbb{R}, 
    \alpha_1 > 0, 
    \beta > 0, 
    \boldsymbol{\epsilon} \in \mathbb{R}^L
    \bigr\}.
    \label{eq:paramspace}
    \end{equation} 
    
    The likelihood contribution for unit $i$ under Type-I censoring is 
    \begin{equation}
        L_i(\boldsymbol{\theta})
        = f(t_i \mid \rho_i, \beta)^{z_i}\,
          S(t_i \mid \rho_i, \beta)^{1 - z_i}
        = \left[\rho_i\beta
          (\rho_i t_i)^{\beta - 1}\right]^{z_i}
          \exp \left[-(\rho_i t_i)^{\beta}\right].
        \label{eq:likelihood}
    \end{equation}
    and the full likelihood function of $\boldsymbol{\theta}$ given the observed data $\mathcal{D}$ is
    \begin{equation}
        L(\boldsymbol{\theta} \mid \mathcal{D})
        = \prod_{i=1}^{n} L_i(\boldsymbol{\theta})
        = \prod_{i=1}^{n}
        \left[
        \rho_i \beta (\rho_i t_i)^{\beta - 1}
        \right]^{z_i}
        \exp\left[
        -(\rho_i t_i)^\beta
        \right],
        \label{eq:fulllik}
    \end{equation}
    where $\rho_i = \exp(\alpha_0 + \alpha_1 s_i + \epsilon_{l_i})$ depends on $\boldsymbol{\theta}$ through $\boldsymbol{\alpha}$, and $\epsilon_{l_i}$. The corresponding log-likelihood function is given by
   
    \begin{equation}
        \ell(\boldsymbol{\theta} \mid \mathcal{D})
        =
        \sum_{i=1}^{n}
        \left\{
        z_i
        \left[
        {\log( \beta)
        +
        \log( \rho_i)}
        +
        (\beta - 1)\log(\rho_i t_i)
        \right]
        -
        (\rho_i t_i)^\beta
        \right\}.
        \label{eq:fullloglik}
    \end{equation}
    
    The $q$-reliable life at stress $s$, defined as the
    time by which a proportion $(1 - q)$ of units have
    failed, is obtained by solving
    $S(\tau_q \mid \rho(s), \beta) = q$, giving
    \begin{equation}
        \tau_q^l(s)
        = \frac{\bigl[-\log(q)\bigr]^{1/\beta}}{\rho(s)}.
        \label{eq:reliablelife}
    \end{equation}
    The quantity of primary interest is the reliable life at the use stress $s^*$, denoted by $\tau_q(s^*)$, assessed against the
    reliability target $\tau_q^*$. Similarly, the $p$-th quantile of the lifetime distribution at stress $s$ with location $l$, defined as the time by which a proportion $p$ of units have failed, is obtained by solving $S(t_p^l \mid \rho(s), \beta) =
    1 - p$, giving
    \begin{equation}
        t_p^l(s)
        = \frac{\bigl[-\log(1-p)\bigr]^{1/\beta}}{\rho(s)}.
        \label{eq:quantile}
    \end{equation}
    Note that the two quantities are related by $t_p^{rb}(s) = \tau_{1-p}(s)$, where the term $rb$ in the superscript is used to represent the reference batch.
      	 
	\subsection{Prior Elicitation}
	\label{sec:prior}

    In reliability and industrial engineering, lifetime characteristics such as quantiles, median life, and mean time to failure at given stress conditions are often more accessible to engineering practitioners than the underlying model parameters. Following this observation, the hyperparameters of the Weibull-Eyring ALT model are determined through a structured prior elicitation procedure described by \cite{OHagan2006} in which expert judgements are sought on observable lifetime characteristics rather than on model parameters directly.
     
    Recall from Eq. \eqref{eq:paramvec} that the parameter vector is $\boldsymbol{\theta} = (\boldsymbol{\alpha}^\top, \beta, \boldsymbol{\epsilon}^\top)^\top \in \Theta$,
    where $\boldsymbol{\alpha} = (\alpha_0, \alpha_1)^\top$ with $\alpha_0 \in \mathbb{R}$ and $\alpha_1 > 0$, $\beta > 0$, and $\boldsymbol{\epsilon} = (\epsilon_1,
    \ldots, \epsilon_L)^\top \in \mathbb{R}^L$. The support constraint on $\beta$ motivates a Gamma prior for $\beta$ and a bivariate normal prior for $\boldsymbol{\alpha}$, while the location effects, being unrestricted and independent of one another, are assigned independent zero-mean normal priors with distinct variances. The complete prior specification is
    \begin{align}
        \beta
        &\sim \text{Gamma}(a_\beta, b_\beta),
        \label{eq:priorbeta}                          \\
        (\alpha_0, \alpha_1)^\top
        &\sim \mathcal{N}_2(\boldsymbol{\mu}_\alpha,
          \boldsymbol{\Sigma}_\alpha),
        \label{eq:prioralpha}                         \\
        \epsilon_l
        &\sim \mathcal{N}(0,\, v_l),
        \quad l = 1, 2, \ldots, L,
        \label{eq:priorepsilon}
    \end{align}
    where the hyperparameters $(a_\beta,\, b_\beta,\, \boldsymbol{\mu}_\alpha,\, \boldsymbol{\Sigma}_\alpha,\, v_l)$ are determined through the three elicitation stages described below. These priors are later used to compute the Bayesian assurance, that is, the probability that the reliability requirement will be met before data
	are collected. The hyperparameters are elicited in three stages, each based on expert assessments of observable lifetime characteristics at specified stress conditions, as described below. 
    
    \textbf{Weibull Shape Parameter.} Rather than eliciting $\beta$ directly, expert judgement is obtained through the ratio of any two distinct lifetime quantiles coming from the same production batch at a given stress level. This ratio is more intuitive in engineering contexts and has the advantage of eliminating the rate parameter $\rho(s)$. For a Weibull distribution with rate parameter $\rho(s)$ and shape parameter $\beta$, the $p$-th quantile of the failure time at stress $s$ and location $l$ is $t_p^l(s) = [-\log(1-p)]^{1/\beta}/\rho(s)$.
    Taking the ratio of the $p_1$-th and $p_2$-th quantiles for $0 < p_1 < p_2 < 1$ eliminates $\rho(s)$, giving
    \begin{equation}
    r_1 = \frac{t_{p_1}^l}{t_{p_2}^l}
        =
        \left(
        \frac{\log(1 - p_1)}{\log(1 - p_2)}
        \right)^{1/\beta}.
        \label{eq:ratio}
    \end{equation}
    Rearranging Eq. \eqref{eq:ratio} expresses $\beta$ as a monotonic decreasing function of $r_1$,
    \begin{equation}
        \beta
        = \frac{\log \left[\log(1-p_1)/\log(1-p_2)
          \right]}{\log(r_1)} = \frac{c_1}{\log(r_1)},
        \label{eq:beta_from_r}
    \end{equation}
    where $c_1 = \log[\log(1-p_1)/\log(1-p_2)] $ is a constant. The expert is therefore asked to assess the ratio $r = t_{p_1}/t_{p_2}$ at a given stress level, where $t_{p}$ is the time by which $p$-th proportion of the units have failed. 
    The expert provides lower, median, and upper quartile assessments of $r_1$, denoted by $r_{1,0.25}, ~ r_{1,0.50}$ and $r_{1,0.75}$, respectively. 
    % \[ 
    % \Pr (r_1 < r_{1,0.25}) = 0.25, \quad \Pr (r_1 < r_{1,0.50}) = 0.50, \text{ and} \quad \Pr (r_1 < r_{1,0.75}) = 0.75.
    % \]

    Since $\beta$ is a monotonic decreasing function of $r_1$ using Eq. \eqref{eq:beta_from_r}, each expert quartile of $r_1$ maps to a corresponding implied quartile of $\beta$, with the ordering reversed:
    \[
        \beta_{0.75}
        = \frac{c_1}{\log(r_{1,0.25})},
        \quad
        \beta_{0.50}
        = \frac{c_1}{\log(r_{1,0.50})}, \text{ and}
        \quad
        \beta_{0.25}
        = \frac{c_1}{\log(r_{1,0.75})},
        \label{eq:beta_quartiles}
    \]
     The hyperparameters $a_\beta$ and $b_\beta$ are then determined to approximately match the $\text{Gamma}(a_\beta, b_\beta)$ distribution with these three quartiles of $\beta$, completing the prior elicitation for the shape parameter.
    
	  \textbf{Stress-life regression parameters.} Recall from Eq. \eqref{eq:prioralpha}, $(\alpha_0, \alpha_1)^\top  \sim \mathcal{N}_2(\boldsymbol{\mu}_\alpha, \boldsymbol{\Sigma}_\alpha)$, for hyper-mean vector $\mu_\alpha = (\bar{\alpha_0},\bar{\alpha_1})^\top$ and variance matrix
      \[
      \Sigma_\alpha
         = 
        \begin{pmatrix}
            \sigma^2_{00} & \sigma^2_{01} \\[4pt]
            \sigma^2_{01} & \sigma^2_{11}
        \end{pmatrix}.
      \]
    Suppose $M_{s}^{l}$ represents the median lifetime at stress level $s$ from location $l$. The prior mean of $\alpha_1$ is determined from expert assessments of the median lifetime at two distinct stress levels coming from same location say $\epsilon_l$; $s_i$, $s_j$ where $i, j \in \{1,2,\ldots,K\}$, \text{and} $i \neq j.$ On the log scale, using Eq. \eqref{eq:quantile}, these medians satisfy
	\begin{align}
        \log(M_{s_m}^{l})
        =
        -\alpha_0-\alpha_1s_m-\epsilon_l
        +\frac{1}{\beta}\log\left(-\log(0.5)\right),
        \qquad m\in\{i,j\}.
		% \log(M_{s_i}^{l}) =  & - \alpha_0 - \alpha_1 s_i - \epsilon_{l} + \frac{1}{\beta} \log(-\log(0.5)), \\
		% \log(M_{s_j}^{l})  = & - \alpha_0 - \alpha_1 s_j - \epsilon_{l} + \frac{1}{\beta} \log(-\log(0.5)), 
	\end{align}  
    By subtracting these equations, we get 
    \begin{align}
    \label{eq_prior_mean_alpha_1}
        \bar{\alpha}_1 =\frac{\log (M_{s_i}^{l} /M_{s_j}^{l})}{ s_j-s_i } = \frac{\log(r_2)}{s_j-s_i},
    \end{align} where $r_2 = M_{s_i}^{l} /M_{s_j}^{l}$; the ratio of median lifetimes at the two stress levels (the quantity the expert is asked to assess). Note that $\epsilon_l = 0$ for a reference batch. Therefore, for a given $\alpha_1$, the prior mean of $\alpha_0$ can be calculated using the $\left(1-e^{-1}\right)$-th quantile from the reference batch at any stress level, say $s_i$, as follows
    \begin{align}
        \bar{\alpha}_0 = -\log \left(t^{rb}_{1-e^{-1}}(s_i)\right)  - \alpha_1\, s_i,
    \end{align}
    where $\alpha_1$ can be replaced by the prior mean value obtained in Eq. \eqref{eq_prior_mean_alpha_1}.  % ********* The expert gives 1- e... which is 0.632 which an equivalent of 2/3rd of the units. 

    Having determined the prior means $\bar{\alpha_0}$ and $\bar{\alpha}_1$ from expert
    assessments, the remaining task is to specify the prior covariance matrix by determining the diagonal elements $\sigma^2_{00}$ and $\sigma^2_{11}$ and the covariance term $\sigma^2_{01}$, which capture marginal prior uncertainty in $\alpha_0$ and $\alpha_1$ individually, together with their prior dependence. Following the approach of \citet{wilsonandfarrow2021}, the three unknown entries $\sigma^2_{00}$, $\sigma^2_{11}$, and $\sigma^2_{01}$ are recovered from expert uncertainty assessments about the log median lifetime at three distinct stress levels, without requiring the expert to reason about model parameters directly.
    
    Consider the $\left(1-e^{-1}\right)$-th quantile from the reference batch at a chosen stress level, $s$, which is given by 
    \begin{align}
       \log \left(t^{rb}_{1-e^{-1}}(s)\right)  = -  \alpha_0 - \alpha_1\, s,
    \end{align} 
    where the only sources of prior uncertainty in $t^{rb}_{1-e^{-1}}(s_i)$ are $\alpha_0$ and $\alpha_1$. Under the bivariate normal prior $(\alpha_0, \alpha_1)^\top \sim \mathcal{N}_2(\boldsymbol{\mu}_\alpha, \boldsymbol{\Sigma}_\alpha)$ specified in Eqs.~(15)--(16), the prior variance of this quantile lifetime at stress $s$ is
    \begin{equation}
        \mathrm{Var} \left[ \log \left(t^{rb}_{1-e^{-1}}(s)\right) \right]   =   \sigma^2_{00} + \sigma^2_{11} s^2 + 2 \sigma^2_{01} s.
    \label{eq:var_decomp}
    \end{equation}
    $\sigma^2_{00}$ captures prior
    uncertainty in the intercept $\alpha_0$; the term $\sigma^2_{11}s^2$ captures
    uncertainty in the slope $\alpha_1$ amplified quadratically by stress; and the cross
    term $2\sigma^2_{01}s$ reflects the prior dependence between $\alpha_0$ and
    $\alpha_1$. Denote the elicited variance quantity by $\delta^2(s)$ which may be constructed from the expert's quartile assessments. The variance in Eq. \eqref{eq:var_decomp} is a quadratic function of stress $s$, meaning that assessments at three distinct stress levels, say $s_i,~s_j,$ and $s_k$ suffice to identify all three unknowns. Provided the stress levels are distinct,  
    \[ %begin{equation}
        \begin{pmatrix}
            \sigma^2_{00} \\[4pt]
            \sigma^2_{11} \\[4pt]
            \sigma^2_{01}
        \end{pmatrix}
         = 
        \mathbf{C}^{-1}
        \begin{pmatrix}
            \delta^2(s_i) \\[4pt] \delta^2(s_j) \\[4pt] \delta^2(s_k)
        \end{pmatrix}, \text{~with~} \mathbf{C}
         = 
        \begin{pmatrix}
            1 & s_i^2 & 2s_i \\[4pt]
            1 & s_j^2 & 2s_j \\[4pt]
            1 & s_k^2 & 2s_k
        \end{pmatrix},
        \label{eq:solution}
    \] where $\mathbf{C}$ is invertible, completes the specification of $\Sigma_\alpha$. 
    
    \textbf{Location Effects.}
    The variance $v_l$ of each location effect $\epsilon_l$
    is elicited by asking the expert to assess the ratio
    of a lifetime quantile at location $l$ compared to
    the reference batch, at the same stress $s$ and
    quantile level $p$. The log-ratio of quantile
    lifetimes satisfies
    \begin{equation}
        \log (R_l)
        = \log \left(
          \frac{t_p^{l}(s)}{t_p^{\mathrm{rb}}(s)}
          \right)
        = -\epsilon_l,
        \label{eq:loc_ratio}
    \end{equation}
    since the reference batch has $\epsilon_l = 0$ by
    convention. Under the prior $\epsilon_l \sim
    \mathcal{N}(0, v_l)$, the log-ratio
    $\log (R_l) \sim \mathcal{N}(0, v_l)$. The expert
    is therefore asked to provide lower and upper
    quartile assessments of $R_l$, denoted
    $r_{l,0.25}$ and $r_{l,0.75}$, satisfying
    \[
        \Pr (R_l < r_{l,0.25}) = 0.25
        \quad \text{and} \quad
        \Pr (R_l < r_{l,0.75}) = 0.75.
    \]
    From the standard normal quartile $z_{0.75} = 0.6745$,
    the implied variance is
    \begin{equation}
        v_l
        = \left(
          \frac{\log(r_{l,0.75}/r_{l,0.25})}{1.349}
          \right)^{ 2},
        \label{eq:vl}
    \end{equation}
    and this is repeated independently for
    each location $l = 1, \ldots, L$.

    As a special case, when a common variance $v_l = v_\epsilon$
    is assumed across all locations and the expert
    compares two arbitrary locations $l$ and $l^*$
    rather than against a reference batch, the
    log-ratio satisfies
    \[
        \log(R_{l,l^*}) = \log \left(
        \frac{t_p^{l}(s)}{t_p^{l^*}(s)}
        \right)
        = \epsilon_{l^*} - \epsilon_l
        \sim \mathcal{N}(0,\, 2v_\epsilon).
    \]
    Denoting the
    expert's lower and upper quartile assessments of
    $R_{l,l^*}$ by $r_{0.25}$ and $r_{0.75}$ respectively,
    the same argument as above gives
    \begin{equation}
        v_\epsilon = \frac{1}{2}
        \left(
        \frac{\log(r_{0.75}/r_{0.25})}{1.349}
        \right)^{ 2},
        \label{eq:vl_kevin}
    \end{equation}
    which coincides with the prior specification of \citet{wilsonandfarrow2021}.

    The entire elicitation is thus grounded in observable lifetime quantities at specified stress conditions, consistent with the guiding principle of the framework articulated in \citet{wilsonandfarrow2021}.
    
	\section{Bayesian Assurance}
    \label{sec:bayesian_assurance}
     
    Bayesian assurance is the pre-experimental probability, evaluated under the manufacturer's design prior
    $p(\boldsymbol{\theta})$, that the planned reliability demonstration test will satisfy the specified posterior decision criterion. The design prior $p(\boldsymbol{\theta})$ represents the manufacturer’s current state of knowledge about the true reliability characteristics of the product, based on engineering judgment and available historical data. The posterior decision criterion requires that the posterior probability of meeting the reliability target exceeds a specified threshold $1 - \delta$, where $\delta \in (0,1)$ is the decision risk, typically $\delta = 0.05$. 
     Whether the test will yield sufficient evidence to meet the reliability requirement is unknown at the design stage.
     
    Formally, 
    \begin{equation}
      % \text{Assurance}
      \mathcal{A} 
       = 
      \int_{\Theta}
        \Pr \bigl(\text{$\mathcal{H}$} \mid \boldsymbol{\theta}\bigr)\,
        p(\boldsymbol{\theta})\,
      d\boldsymbol{\theta},
      \label{eq:assurance_definition}
    \end{equation}
    where $\mathcal{A}$ is the Assurance, $\mathcal{H}$ represents Test Passed and $\Pr \bigl(\text{$\mathcal{H}$}\mid \boldsymbol{\theta}\bigr)$ is the  conditional probability of a successful test outcome given the true parameters.  
    The test is declared passed if the posterior probability satisfies
    \begin{equation}
      \Pr \bigl(\tau_{q}(s^{*}) \geq \tau_{q}^{*}
                     \mid \mathcal{D}\bigr)
       \geq 
      1 - \delta,
      \label{eq:decision_criterion}
    \end{equation}
    
    Substituting \eqref{eq:decision_criterion} into \eqref{eq:assurance_definition},
    \begin{equation}
        \text{$\mathcal{A}$} = \Pr \left(\Pr \left(\tau_q(s^*) \geq \tau_q^* \mid \mathcal{D}\right) \geq 1 - \delta\right),
        \label{eq:assurance_full}
    \end{equation}
    where the outer probability is taken jointly over $p(\boldsymbol{\theta})$ and the induced predictive distribution over $\mathcal{D}$. The elicited prior distributions introduced in Section \ref{sec:prior} provide the basis for deriving a tractable lower bound on the quantity of interest, which is the design prior probability that the reliable life exceeds the target before any data are observed providing an upper bound on the assurance achievable for any finite sample size.

    \subsection{Monte Carlo Estimation of the Assurance Curve}
    \label{subsec:assurance_mc}
    
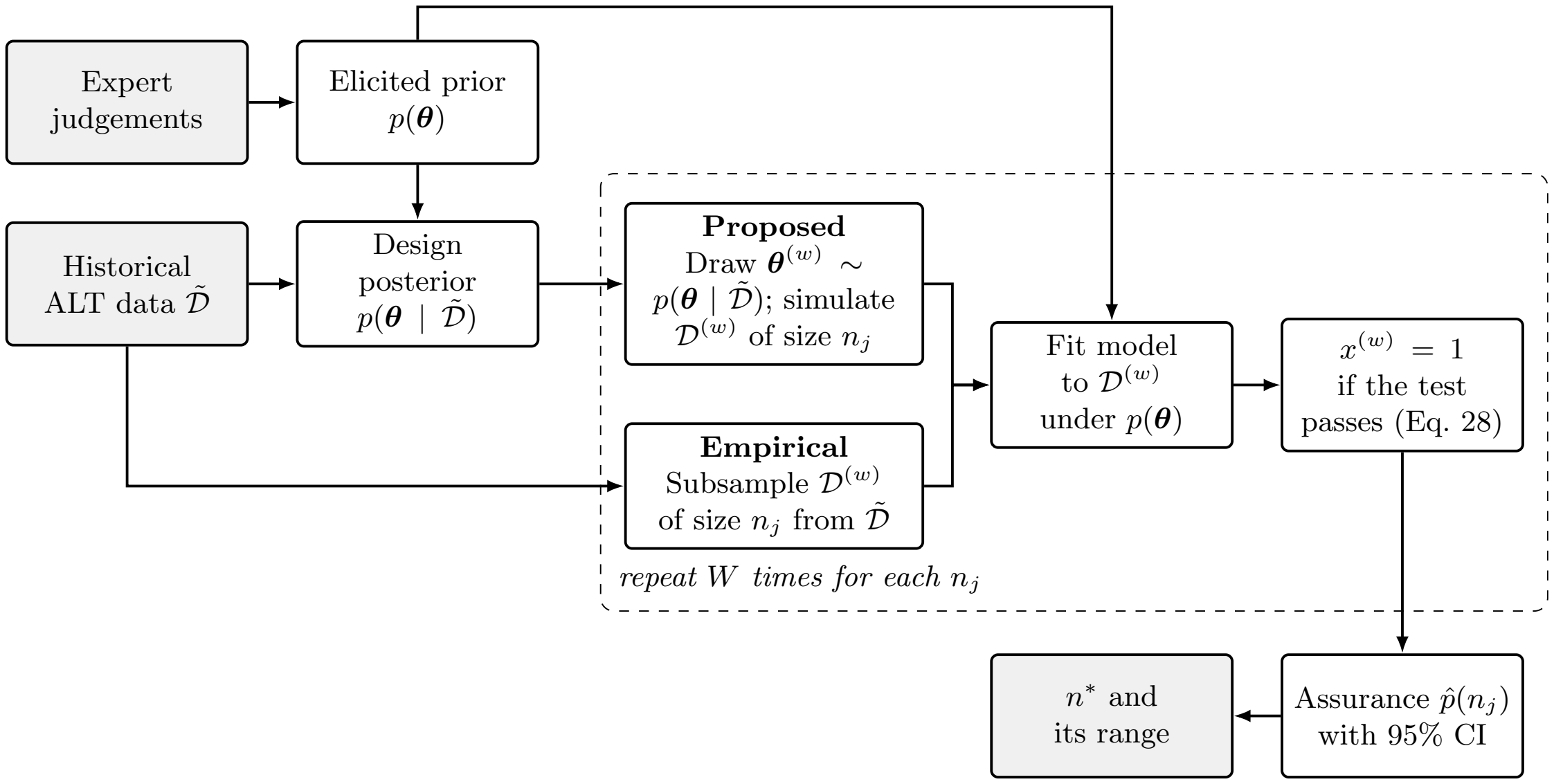
\begin{figure}[!t]
        \centering
        \resizebox{0.9\textwidth}{!}{\input{flowchart_tikz}}
        \caption{Overview of the assurance procedure under the proposed and empirical approaches.}
        \label{fig:flowchart}
    \end{figure}

    An assurance curve characterises how the probability of meeting a predefined reliability criterion varies with the number of units included in a test. This enables the assessment of how uncertainty in the model parameters affects the likelihood of a successful reliability demonstration and provides a basis for evaluating and selecting appropriate test designs.
    The procedure for estimating the assurance curve is summarised in Figure~\ref{fig:flowchart}. For each candidate sample size, the test data in each replication are obtained either by simulation from the design posterior (the proposed approach) or by subsampling the historical data (the empirical approach, used for comparison in Section~\ref{subsec:assurance_results}); the remaining steps are common to both.
    
    The assurance in Eq. \eqref{eq:assurance_full} is estimated at each candidate sample size $n_j$ by computing the proportion of simulated test replications that satisfy the posterior decision criterion in Eq. \eqref{eq:decision_criterion}. In contrast to the curve-fitting scheme of \citet{Muller1999} and \citet{muller1995}, the assurance is estimated separately at each of candidate sample sizes, and the resulting assurance values are plotted as a function of $n$, from which the minimum sample size meeting the target assurance $\gamma$ is identified. The following six steps consist of the procedure for determining the sample size that meets the desired assurance for a given target reliable lifetime value. 
    
    \medskip
    \noindent\textbf{Step 1: Design posterior.} Fit the model to the historical dataset $\tilde{\mathcal{D}}$ via MCMC under the design prior $p(\boldsymbol{\theta})$ to obtain draws $\{\tilde{\boldsymbol{\theta}}^{(m)}\}_{m=1}^{M_0}$ from the design posterior $p(\boldsymbol{\theta} \mid \tilde{\mathcal{D}})$. 

    \medskip
    \noindent\textbf{Step 2: Design grid.} Select $M$ candidate sample sizes $n_j$, together with the associated test stress allocation $\mathbf{s}_{\text{test}} = (s_{\text{test},1}, \ldots, s_{\text{test},n_j})$, with each in the permissible stress set $\{s_1, \ldots, s_K\} \subset (s_0, s_h]$. 
    
    \medskip
    \noindent\textbf{Step 3: Simulate data from the design posterior.} For each $n_j$, draw a parameter value $\boldsymbol{\theta}^{(w)} = \bigl(\beta^{(w)}, \alpha_0^{(w)}, \alpha_1^{(w)}, \boldsymbol{\epsilon}^{(w)}\bigr)$ at random from the design posterior of Step 1, and simulate failure times
    \begin{equation*}
    	T_i^{(w)} \sim \mathrm{Weibull}\bigl(\rho_i^{(w)}, \beta^{(w)}\bigr),
    	\qquad
    	\rho_i^{(w)} = \exp\bigl(\alpha_0^{(w)} + \alpha_1^{(w)} s_i + \epsilon_{l_i}^{(w)}\bigr),
    	\qquad i = 1, \ldots, n_j,
    \end{equation*}
    where $s_i$ and $l_i$ are the stress level and location assigned to unit $i$ in Step 2. Applying Type-I censoring at $t_C$ gives $t_i^{(w)} = \min(T_i^{(w)}, t_C)$ and $z_i^{(w)} = \mathbf{1}\{T_i^{(w)} \leq t_C\}$, forming the simulated dataset $\mathcal{D}^{(w)} = \{(t_i^{(w)}, z_i^{(w)}) : i = 1, \ldots, n_j\}$.
  
    \medskip
    \noindent\textbf{Step 4: Evaluate the pass/fail indicator.} For each {$\tilde{\mathcal{D}}^{(w)}$}, evaluate the model via MCMC under ${p(\boldsymbol{\theta})}$ and obtain posterior draws $\{\boldsymbol{\theta}^{(b)}\}_{b=1}^{B}$. Compute the binary indicator
    \begin{equation}
        {x^{(w)}} = \mathbf{1} \left[\Pr \left(\tau_q(s^*) \geq \tau_q^* \mid {\tilde{\mathcal{D}}^{(w)}}\right) \geq 1 - \delta\right].
        \label{eq:pass_fail}
    \end{equation}
    
    \medskip
    \noindent\textbf{Step 5: Estimate the empirical assurance estimate.} Repeat Steps 2--3 independently $W$ times at each $n_j$ to obtain the Monte Carlo assurance estimate
    \begin{equation}
        \hat{p}(n_j) = \frac{1}{W}\sum_{w=1}^W x^{{(w)}} \xrightarrow{W \to \infty} \Pr (\text{$\mathcal{H}$} \mid n = n_j).
        \label{eq:mc_estimate}
    \end{equation}
    
    \medskip
    \noindent\textbf{Step 6: {Determine $n^*$.}}
    Plot the assurance estimates $\hat{p}(n_j)$ against $n_j$ for $j = 1, \ldots, M$. The minimum sufficient sample size at assurance target $\gamma$ is then
    \begin{equation}
        n^* = \min \left\{ n \in \mathbb{Z}_{>0} : {\hat{p}(n)}  \geq \gamma\right\},
        \label{eq:nstar}
    \end{equation}
    where {$\mathbb{Z}_{>0}$ denotes a set of positive natural numbers} and $\gamma \in (0,1)$ is typically set to $0.80$ or $0.90$ in practice. 

    {The complete procedure is summarized in Algorithm \ref{alg:assurance1}.} To quantify the Monte Carlo uncertainty in the estimated assurance at each candidate sample size, note that each $x^{(w)}$ is an independent Bernoulli trial with success probability $p(n_j)$. The estimator $\hat{p}(n_j)$ therefore has standard error
    \begin{equation}
        \text{SE} \bigl(\hat{p}(n_j)\bigr)
        = \sqrt{\frac{\hat{p}(n_j)
          \bigl(1 - \hat{p}(n_j)\bigr)}{W}},
        \label{eq:se}
    \end{equation}
    and the approximate 
    $(1-\eta) 100 $\% confidence interval for $p(n_j)$ is
    \begin{equation}
        \text{CI}_{1-\eta} \bigl(
        \hat{p}(n_j)\bigr)
        = \hat{p}(n_j)  \pm 
          z_{\eta/2}\,\text{SE} \bigl(\hat{p}(n_j)\bigr),
        \label{eq:ci}
    \end{equation}
    where $z_{\eta/2}$ is the upper $\eta/2$ quantile of the standard normal distribution. The confidence band is plotted alongside the assurance curve in Section \ref{subsec:assurance_results}, providing a visual measure of Monte Carlo uncertainty at each $n_j$ and at the proposed $n^*$.

    \begin{algorithm}[H]
    \caption{Bayesian assurance estimation}
    \label{alg:assurance1}
    \begin{algorithmic}[1]
    \Require Observed data $\tilde{\mathcal{D}}$;
    candidate sizes $n_1, \ldots, n_M$;
    replications $W$; decision risk $\delta$;
    assurance target $\gamma$
    \Ensure $\hat{p}(n_j)$ for each $n_j$,
    $\mathrm{SE}(\hat{p}(n_j))$, 95\% CI for each $n_j$;
  $n^{*}$ and its range $[n^{*}_{\mathrm{L}}, n^{*}_{\mathrm{U}}]$ %on $\hat{p}(n_j)$, and $n^*$
    \State Fit model to
$\tilde{D}$ using MCMC under $p(\boldsymbol{\theta})$; retain draws
$\{\tilde{\boldsymbol{\theta}}^{(m)}\}_{m=1}^{M_0}$ from the design posterior
$p(\boldsymbol{\theta} \mid \tilde{D})$
    \For{$j = 1$ to $M$}
        \For{$w = 1$ to $W$}
            \State Draw $\bm{\theta}^{(w)}$ at random from $\{\tilde{\bm{\theta}}^{(m)}\}$
            \State Simulate $\mathcal{D}^{(w)}$ of size $n_j$ from the Weibull--Eyring model with $\bm{\theta}^{(w)}$, censored at $t_C$
            \State Fit model to $\mathcal{D}^{(w)}$
            via MCMC under $p(\boldsymbol{\theta})$;
            retain posterior draws
            $\{\boldsymbol{\theta}^{(b)}\}_{b=1}^{B}$
            \State Compute posterior pass probability:
            \[
                \hat{p}_{w} = \frac{1}{B}
                \sum_{b=1}^{B}
                \mathbf{1}\bigl\{
                \tau_q^{(b)}(s^*) \geq \tau_q^*
                \bigr\}
            \]
            \State Set pass indicator:
            $x^{(w)} = \mathbf{1}\bigl\{
            \hat{p}_{w} \geq 1 - \delta \bigr\}$
        \EndFor
        \State $\hat{p}(n_j) \gets
        \dfrac{1}{W} \sum_{w=1}^{W} x^{(w)}$
        \State $\mathrm{SE}(\hat{p}(n_j)) \gets
        \sqrt{\hat{p}(n_j)
        \bigl(1 - \hat{p}(n_j)\bigr) / W}$
        % \State Compute $95$\% CI on $\hat{p}(n_j)$
        \State $ \text{CI}_{0.95} \bigl(
        \hat{p}(n_j)\bigr) \gets \hat{p}(n_j)  \pm 
          1.96 \,\text{SE} \bigl(\hat{p}(n_j)\bigr)$
    \EndFor
    \State $n^{*} \gets \lceil n_{\gamma} \rceil$, where $n_{\gamma}$ is the point at which the curve joining $\{(n_j, \hat{p}(n_j))\}$ first reaches $\gamma$ 
    \State { $n^{*}_{\mathrm{L}}, n^{*}_{\mathrm{U}} \gets$ the same, for the upper and lower CI limits}
    \end{algorithmic}
    \end{algorithm}

    \subsection{Incorporating Historical Data}
    \label{subsec:historical}
    
    In industrial settings where failure data from prior product iterations are available, it would be inefficient to discard such evidence at the test planning stage. Suppose a historical dataset $\tilde{\mathcal{D}}$ of size $n_0$ is available, collected under the same Weibull--Eyring model as specified in Section \ref{sec:model}; in the present study, $\tilde{\mathcal{D}}$ consists of the $n_0 = 90$ SCBA observations described in Section~\ref{sec:industrial_background}. In the proposed approach, fitting the model to $\tilde{\mathcal{D}}$ via MCMC yields the design posterior $p(\boldsymbol{\theta} \mid  \tilde{\mathcal{D}})$, from which the parameter values $\boldsymbol{\theta}^{(w)}$ in Step~3 of Subsection~\ref{subsec:assurance_mc} are drawn. When $\tilde{\mathcal{D}}$ is consistent with $\tau_q(s^*) \geq \tau_q^*$, the design posterior concentrates in the favourable region of parameter space, yielding a higher assurance for any fixed $n$ and potentially reducing the required sample size to achieve target $\gamma$.
    
    \textit{Large historical datasets.} When $\tilde{\mathcal{D}}$ is large and representative of the units to be tested, the historical data can also be used directly to estimate the assurance empirically from the data alone, without fitting the model first. In this empirical approach (Figure~\ref{fig:flowchart}), 
    Step~1 is omitted, and the simulation in Step~3 is replaced by resampling the observed units.
    For each $n_j$, draw a subsample of size $n_j$ without replacement from the historical dataset $\tilde{\mathcal{D}}$, according to a prespecified allocation across stress levels. The subsample is denoted $\mathcal{D}^{(w)} = \{(t_i^{(w)}, z_i^{(w)}) : i = 1, \ldots, n_j\}$. Because every subsample comes from the same finite set of units, this approach treats the observed units as the true population and ignores uncertainty about $\boldsymbol{\theta}$, and therefore tends to give a smaller $n^*$. The two approaches are compared in Section~\ref{subsec:assurance_results}.
    
    \textit{No historical data.} It is worth noting that historical data of this kind may not always be available in practice. Constraints on cost, time, or data availability may prevent the collection of a sufficiently large or informative $\tilde{\mathcal{D}}$. In such situations, the proposed methodology remains applicable, with only Step~1 modified: the draws $\{\tilde{\boldsymbol{\theta}}^{(m)}\}_{m=1}^{M_0}$ are generated directly from the design prior instead of the design posterior. 
    Consequently, at each repetition $w$ of Step~3, true parameter values are drawn from the design prior,
    \begin{equation}
        \boldsymbol{\theta}^{(w)} = \bigl(\beta^{(w)}, \alpha_0^{(w)}, \alpha_1^{(w)}, \boldsymbol{\epsilon}^{(w)}\bigr) \sim p(\boldsymbol{\theta}),
        \label{eq:prior_draw}
    \end{equation}
    rather than from $p(\boldsymbol{\theta} \mid \tilde{\mathcal{D}})$, and the remaining steps are unchanged.
    Each draw $\boldsymbol{\theta}^{(w)}$ represents one plausible true state of nature, and averaging the pass indicator $x^{(w)}$ over $W$ such draws approximates the outer integral in the assurance definition Eq. \eqref{eq:assurance_full}. In this setting, the choice of $n^*$ becomes more sensitive to the elicited prior specification, and the incorporation of a utility-based cost framework to guide the sample size decision is a natural extension, left as a direction for future research.
      
	\section{Simulation Study}
	   \label{sec:simulations}
    This section presents the results of the simulation study conducted to determine the minimum sample size $n^*$ required for the proposed Bayesian reliability demonstration test. The study proceeds in three stages: prior elicitation from two engineering experts (Section \ref{subsec:prior_elicitation}), assessment of MCMC convergence across all candidate sample sizes (Section \ref{subsec:converg_diagno}), and construction of the assurance curve and identification of $n^*$ (Section \ref{subsec:assurance_results}).
    
	\subsection{Prior Elicitation Results}
    \label{subsec:prior_elicitation}
    
    \subsubsection{Elicited Expert Judgements}
    \label{subsubsec:elicited-values}
     
    Table \ref{tab:elicited_values} summarises the raw expert assessments collected for each stage of the elicitation procedure, prior to conversion into distributional hyperparameters. The elicitation questions put to the experts are given in Appendix~\ref{sec:elicitation-protocol}. Values are reported on the observable lifetime scale (hrs, or dimensionless ratios) as presented to the expert, consistent with the quantile- and ratio-based elicitation approach adopted in Section \ref{sec:prior}.
    
\begin{table}[htbp]
    \centering
    \scriptsize
    \small
    \setlength{\tabcolsep}{3pt}
    \renewcommand{\arraystretch}{1.2}

    \begin{tabular}{lcccccc}
        \hline
        Quantity & \makecell{Temperature \\ at $s^{raw}$ ($^\circ$C)}
        & Expert & Q1 (0.25) & Median (0.50) & Q3 (0.75) & Other \\
        \hline

        $r_1 = t_{1/3}^{rb}(s)/t_{2/3}^{rb}(s)$ & 60
        & \textbf{1} & 0.54 & 0.60 & 0.66 & -- \\
        & & \textbf{2} & 0.50 & 0.55 & 0.65 & -- \\

        $M_{s_i}^{rb}$ (hrs) & 60
        & \textbf{1} & -- & 140 & -- & -- \\
        & & \textbf{2} & -- & 140 & -- & -- \\

        $M_{s_j}^{rb}$ (hrs) & 72
        & \textbf{1} & -- & 70 & -- & -- \\
        & & \textbf{2} & -- & 70 & -- & -- \\

        $t_{1-e^{-1}}^{rb}(s)$ (hrs) & 60
        & \textbf{1} & -- & -- & -- & 250 \\
        & & \textbf{2} & -- & -- & -- & 250 \\

        \makecell[l]{$\delta^2(s) = \mathrm{Var}[\log(t_{1-e^{-1}}^{rb}(s))]$}
        & \makecell{60, 72, 84}
        & \textbf{1} & -- & -- & -- & 0.18, 0.20, 0.24 \\
        & & \textbf{2} & -- & -- & -- & 0.15, 0.18, 0.25 \\

        $R = t_p^{\text{A}}(s)/t_p^{\text{B}}(s)$ & 60
        & \textbf{1} & 0.65 & -- & 1.50 & -- \\
        & & \textbf{2} & 0.8 & -- & 1.25 & -- \\

        \hline
    \end{tabular}
    \caption{Elicited expert assessments for prior specification of the Bayesian Weibull--Eyring ALT model (Expert 1 and 2).}
    \label{tab:elicited_values}
\end{table}
    
     Quartile assessments for $r_1$ and $R$ were elicited directly from the expert as bounded ratios, whereas the variances $\delta^2(s)$ of the log-scaled $(1-e^{-1})$-th quantile lifetime informing $\Sigma_\alpha$ were assessed directly at three distinct thermal stress levels, thereby ensuring that the covariance structure could be uniquely identified.
     
     These elicited judgements were then propagated through the elicitation equations presented in Section \ref{sec:prior} to derive the hyperparameters governing the prior distributions of $\beta$, $(\alpha_0, \alpha_1)^{\top}$, and $\epsilon_l$.
    
    \subsubsection{Summary of Elicited Hyperparameters}
    \label{subsubsec:hyperparameter-summary}

    Propagating the elicited judgements from section \ref{subsubsec:elicited-values} through the prior elicitation equations yields the full prior specification for the Bayesian Weibull--Eyring ALT model. 
    Table \ref{tab:hyperparameter-summary} summarises the full specification. The shape parameter is %fitted as 
    $\beta \sim \text{Gamma}(a_\beta = 12.021,\ b_\beta =5.896)$. The regression parameters $(\alpha_0, \alpha_1)^{{\top}}$ present in the stress-life relationship are assigned a bivariate normal prior with mean $\boldsymbol{\mu}_\alpha = (14.415,\ -19.259)^{\top}$ and covariance $\Sigma_\alpha$ as presented in Table \ref{tab:hyperparameter-summary}.
    This confirmed {$\Sigma_\alpha$ positive definite (eigenvalues $19.263$ and $0.085$), as required for a valid prior}. The strong negative correlation between $\alpha_0$ and $\alpha_1$ reflects the expected intercept--slope trade-off in the log-linear stress--life relationship, consistent with the reference-stress anchoring used to elicit $\boldsymbol{\mu}_\alpha$. The random location effects are assigned the common variance $v_\epsilon = 0.192$. 
    The corresponding hyperparameters for Expert 2 are given in Table \ref{tab:hyperparameter-summary}, and both priors are shown in Figure \ref{fig:Prior_Posterior}; Expert 2 is less certain about the shape parameter and the stress--life relationship, but more confident that the locations behave alike.
    
    \begin{table}[H]
        \centering
        \small
        \begin{tabular}{llcc}
            \hline
            Expert & Parameter & Prior & Hyperparameter value(s) \\
            \hline
            \multirow{4}{*}{1} & $\beta$ & Gamma$(a_\beta, b_\beta)$ & $a_\beta = 12.021,\ b_\beta = 5.896$ \\
            % $\mathbb{E}[\beta]$ & -- & 2  \\
            & $(\alpha_0, \alpha_1)^{\top}$ & $N_2(\boldsymbol{\mu}_\alpha, \Sigma_\alpha)$ & $\boldsymbol{\mu}_\alpha = (14.415,\ -19.259)^{\top}$ \\
             & &  & $\Sigma_\alpha = \begin{pmatrix} 10.209 & -9.574 \\ -9.574 & 9.139 \end{pmatrix}$ \\
            & $\epsilon_l$ & $N(0, v_\epsilon)$ & $v_\epsilon = 0.192$ \\
            \hline
            \multirow{4}{*}{2} & $\beta$ & Gamma$(a_\beta, b_\beta)$ & $a_\beta = 8.343,\ b_\beta = 4.351$ \\
         & $(\alpha_0, \alpha_1)^\top $ & $N_2(\boldsymbol{\mu}_\alpha, \Sigma_\alpha)$ & $\boldsymbol{\mu}_\alpha = (14.415,\ -19.259)^{\top}$ \\
         & & & ${\Sigma_\alpha =} \begin{pmatrix} 19.622 & -18.717 \\ -18.717 & 17.991 \end{pmatrix}$ \\
         & $\epsilon_l$ & $N(0, v_\epsilon)$ & $v_\epsilon = 0.055$ \\
        \hline
        \end{tabular}
         \caption{Summary of elicited hyperparameters for the Bayesian Weibull--Eyring ALT model (Expert 1 and Expert 2).}
        \label{tab:hyperparameter-summary}
    \end{table}
    \subsection{Posterior Inference and Convergence Diagnostics}
\label{subsec:converg_diagno}

The model is fitted to the $90$ SCBA observations under the prior of each expert separately using JAGS 4.3.1 through the \texttt{rjags} package in \textsf{R}. The expert priors are considered independently rather than combined into a single prior distribution to preserve the distinct prior beliefs and judgements expressed by each expert. This approach provides an assessment on how different expert opinions affect the reliability assurance estimates and model outcomes without forcing those opinions to be weighted.  Each fit used three parallel chains, with $1000$ burn-in iterations followed by $2000$ retained iterations per chain, giving $6000$ posterior draws. Since the assurance procedure of Section \ref{subsec:assurance_mc} repeatedly refits the model to subsamples of the data, convergence was examined for the full dataset and for subsamples of size $n = 6$ and $n = 30$ under each expert's prior. Convergence was equally satisfactory under both priors, so the diagnostics and posterior summaries are reported for Expert 1 only. At all three sample sizes, the Gelman--Rubin statistic $\hat{R}$ lies in $[1.000,1.002]$ and the effective sample sizes (ESS) were observed to be between $[2737,6154]$, indicating that the chains have converged and that the retained posterior draws are adequate for inference. The full diagnostics are given in Appendix \ref{app:convergenceA}.

Figure \ref{fig:Prior_Posterior} compares the prior and posterior densities of the model parameters under the priors of the two experts. For every parameter, the posterior is markedly more concentrated than the prior, most noticeably for $\beta$, $\epsilon_1$ and $\epsilon_2$, showing that the accelerated test data are informative about all the model parameters. The posterior inference is therefore driven mainly by the data and is robust to the choice of expert.   

\begin{figure}[H]
    \centering
    \includegraphics[width=0.85\textwidth]{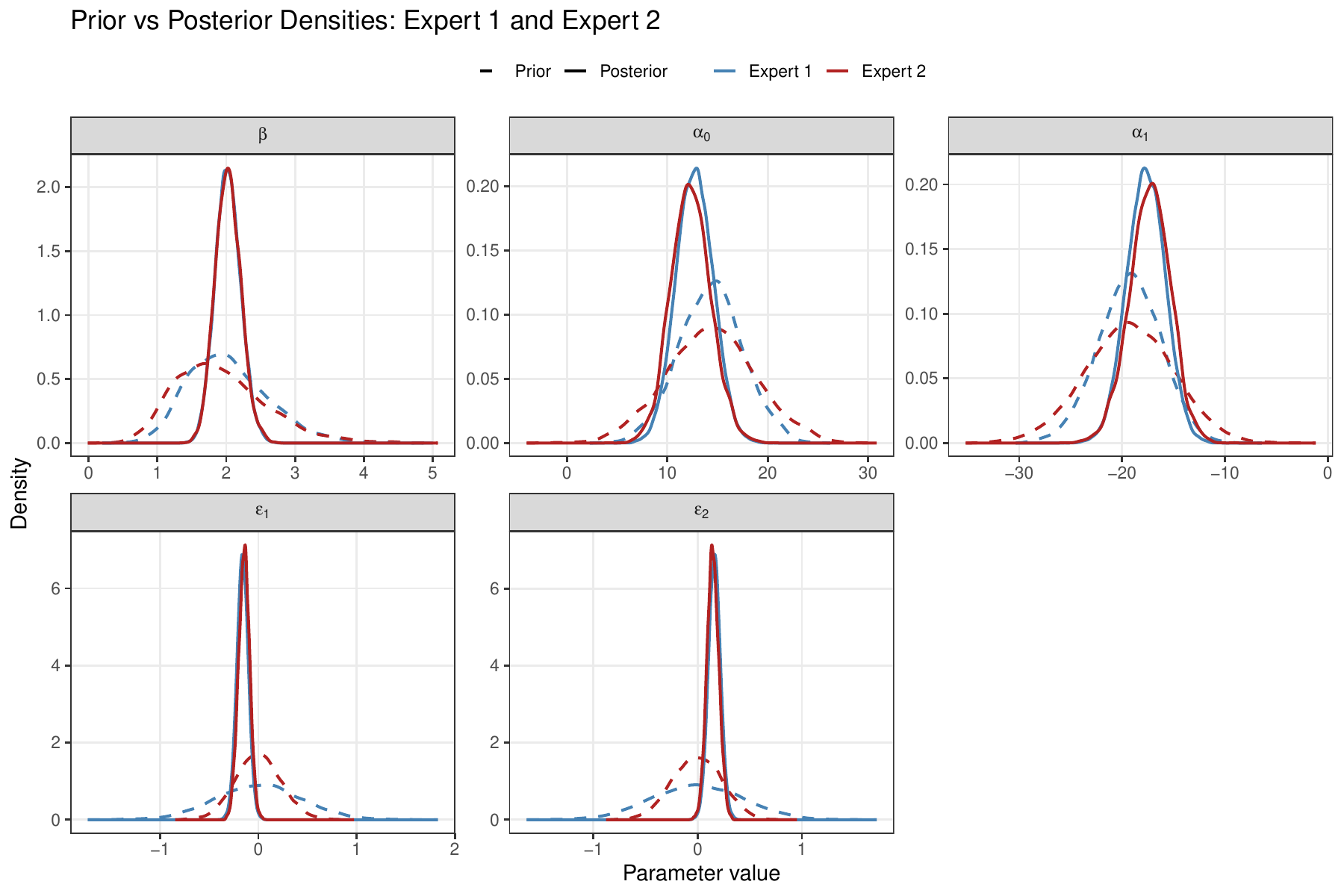}
    \caption{Prior (dashed) and posterior (solid) densities of $\beta$, $\alpha_0$, $\alpha_1$, $\epsilon_1$ and $\epsilon_2$ under the priors of Expert 1 (blue) and Expert 2 (red).}
    \label{fig:Prior_Posterior}
\end{figure}

For the full dataset under the Expert 1 prior (Table \ref{tab:convergence_by_n}), the posterior mean of $\beta$ is $2.02$ with $95\%$ credible interval $(1.68, 2.39)$. Since this interval lies entirely above one, the failure rate increases with time, which is consistent with the thermal degradation mechanisms described in Section \ref{sec:industrial_background}. The posterior mean of $\epsilon_1$ is $-0.16$, with $95\%$ credible interval $(-0.28, -0.04)$, which excludes zero. Since $\epsilon_2 = -\epsilon_1$, the ratio of any lifetime quantile at location A to the same quantile at location B, at any stress level, is $\exp(\epsilon_2 - \epsilon_1)$. At the posterior means, this ratio is approximately $\exp(0.32) \approx 1.38$, so units from location A have systematically longer lifetimes than those from location B. At $n = 6$ and $n = 30$, the credible intervals for $\epsilon_1$, $(-0.73, 0.30)$ and $(-0.46, 0.05)$, include zero, and so do those for $\epsilon_2$. With smaller test plans, the data therefore cannot distinguish between the two locations, and so the location effect makes the predicted lifetimes more uncertain at every stress level, including the use condition at which the reliability requirement is assessed.

    \subsection{Bayesian Assurance and the determination of sample size \texorpdfstring{$n^{*}$}{n*}}
    \label{subsec:assurance_results}
         
    The quantity of interest is the reliable life at $q = 0.90$, the time beyond which 90\% of units survive, and the requirement
    is $\tau_{0.90}^{*} = 50$ hrs at the use condition $s^{*} = 60\,^{\circ}$C ; a less demanding requirement $40$ hrs is also considered. The two experts' elicited hyperparameters (Table~\ref{tab:hyperparameter-summary}) differ in a meaningful way. Expert~2 is less certain about the Weibull shape and the stress--life relationship, but more confident that the locations behave alike.
    
    For each expert, the design posterior is obtained by fitting the model to the $90$ historical units under that expert’s prior, with the MCMC settings of Section \ref{subsec:converg_diagno}. These draws summarise what is believed about the product after the historical test, and they are used to generate plausible outcomes of a future demonstration test. Candidate sample sizes $n_j \in \{6, 12, \ldots, 120\}$ are considered (up to $90$ for the $40$ hrs requirement), adding two units per stress level at each step, with $R = 500$ replications at each $n_j$. 
    Assurance is estimated with the proposed approach (Figure \ref{fig:flowchart}), with $\delta = 0.05$ and $\tau_{0.90}$ evaluated at the reference location. The minimum sample size $n^{*}$ is the smallest integer at which the linearly interpolated assurance curve reaches $\gamma = 0.90$, and the points at which the upper and lower $95\%$ interval limits reach $\gamma$ give a plausible range for $n^{*}$.

    \begin{figure}[!t]
        \centering
        \begin{subfigure}[t]{0.49\textwidth}
            \centering
            \includegraphics[width=\textwidth]{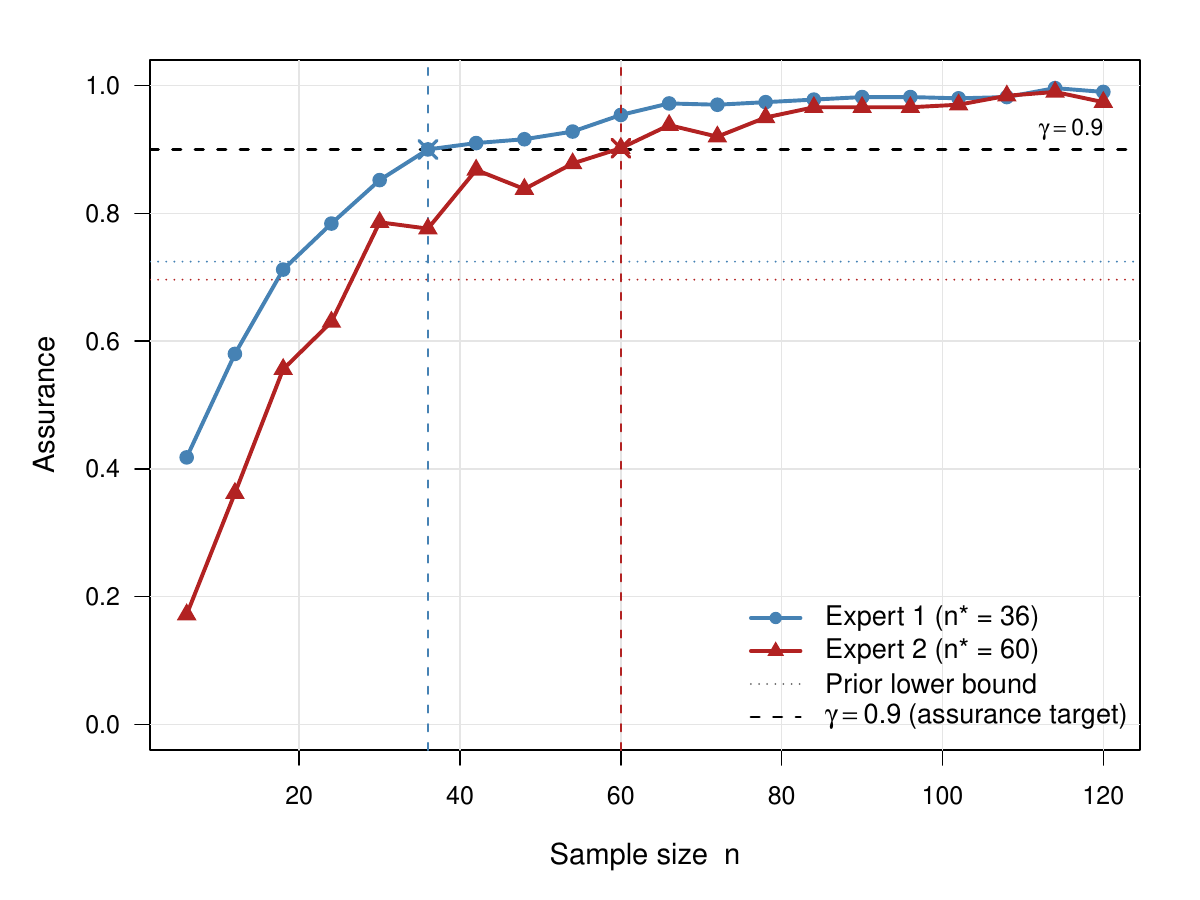}
            \caption{$\tau_q^* = 50$ hrs}
            \label{fig:assurance_tau50}
        \end{subfigure}
        \hfill
        \begin{subfigure}[t]{0.49\textwidth}
            \centering
            \includegraphics[width=\textwidth]{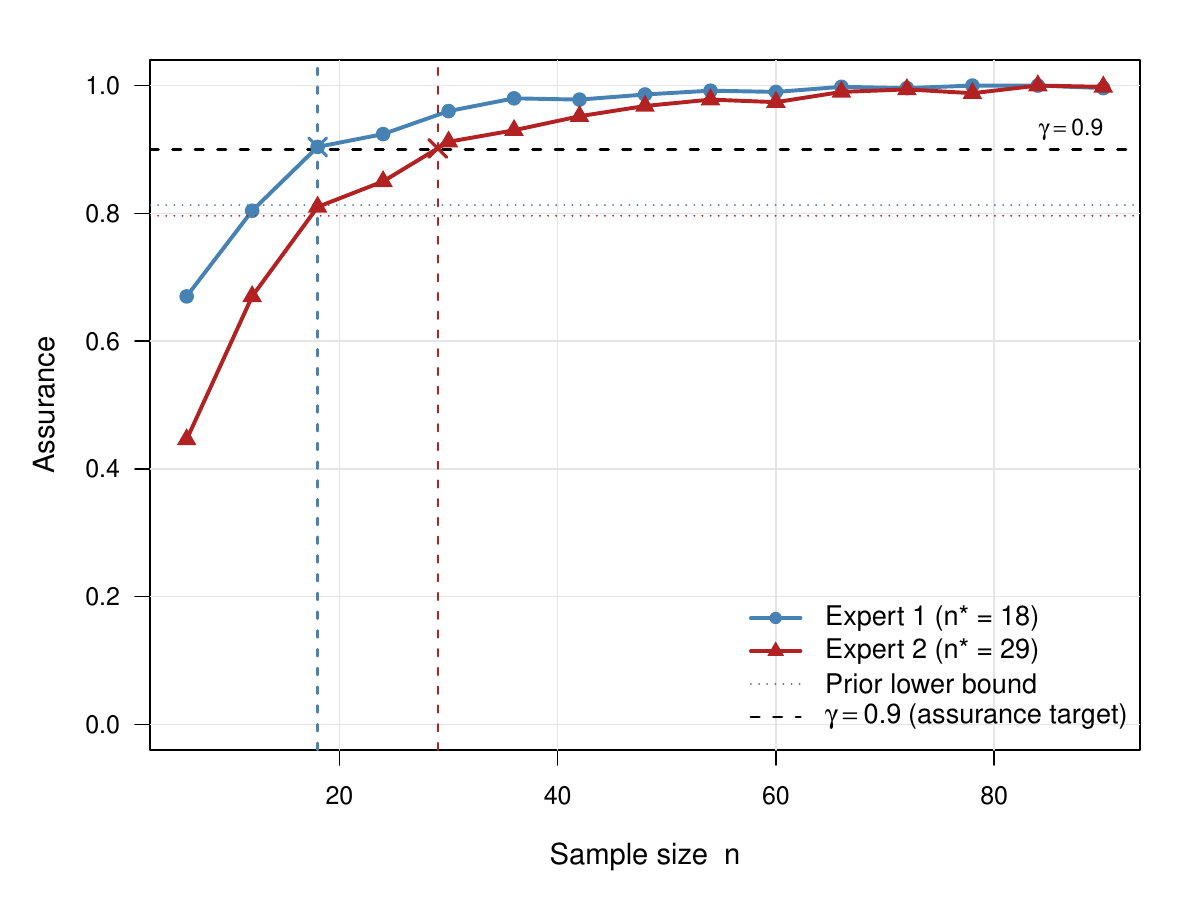}
            \caption{$\tau_q^* = 40$ hrs}
            \label{fig:assurance_tau40}
        \end{subfigure}
        \caption{Assurance curves under the priors of the two experts for target reliable lives of (a) 50 hrs and (b) 40 hrs.}
        \label{fig:assurance_curves}
    \end{figure}    

    Figure~\ref{fig:assurance_curves} overlays the assurance curves of the two experts  for requirements of $\tau_{0.90}^{*} = 50$hrs (Figure~\ref{fig:assurance_tau50}) and $40$ hrs (Figure~\ref{fig:assurance_tau40}). 
    For both experts, assurance increases with $n$, because larger tests estimate the model parameters more precisely and fewer simulated tests fail through parameter uncertainty alone. At $\tau_{0.90}^{*} = 50$ hrs, Expert~1 reaches $\gamma$ at $n^{*} = 36$ and Expert~2 at $n^{*} = 60$. Relaxing the requirement to $\tau_{0.90}^{*} = 40$ hrs reduces these to $n^{*} = 18$ and $n^{*} = 29$, respectively, roughly halving the required test. The larger sample sizes for Expert~2 reflect that expert's greater prior uncertainty, mainly in $(\alpha_0,\alpha_1)$, whose prior variances are almost twice those of Expert~1 (Table~\ref{tab:hyperparameter-summary}).
    
    \begin{figure}[!t]
        \centering
        \begin{subfigure}[t]{0.49\textwidth}
            \centering
            \includegraphics[width=\textwidth]{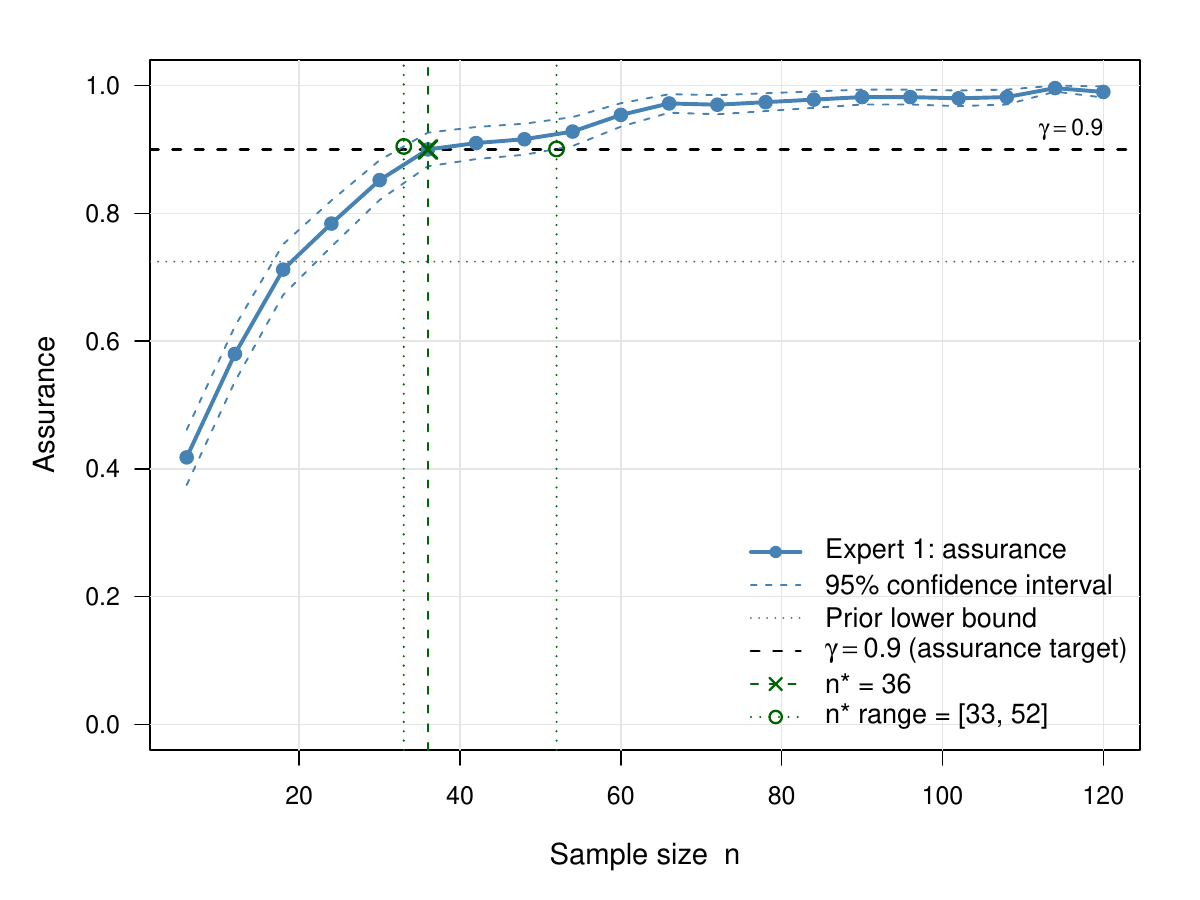}
            \caption{Expert 1}
            \label{fig:assurance_CI_expert1}
        \end{subfigure}
        \hfill
        \begin{subfigure}[t]{0.49\textwidth}
            \centering
            \includegraphics[width=\textwidth]{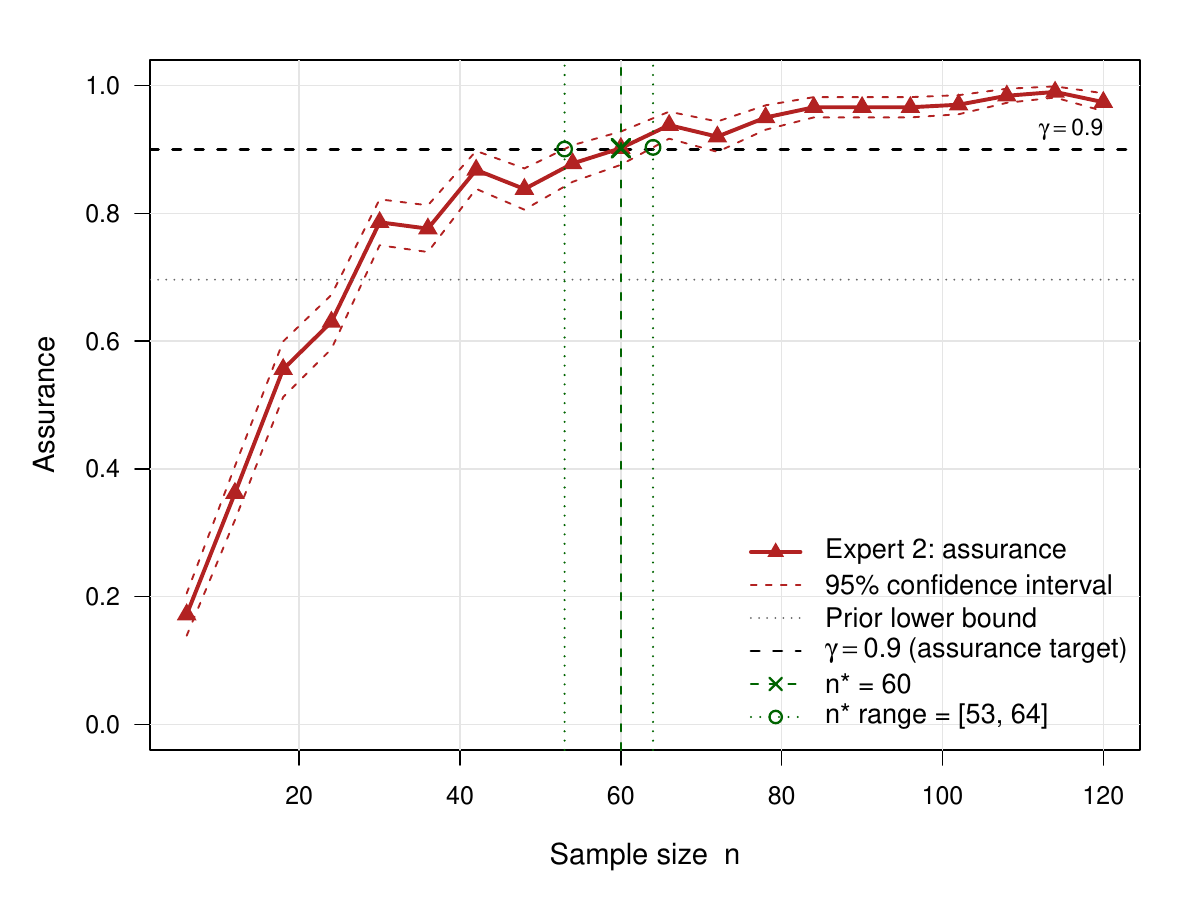}
            \caption{Expert 2}
            \label{fig:assurance_CI_expert2}
        \end{subfigure}
        \caption{Assurance curves with 95\% Monte Carlo confidence intervals for a target reliable life of 50 hrs under the priors of (a) Expert 1 and (b) Expert 2.}
        \label{fig:assurance_CI}
    \end{figure}

    Figure~\ref{fig:assurance_CI} shows each expert's curve separately, with the 95\%
    interval $\hat{p}(n_j) \pm 1.96\,\mathrm{SE}\{\hat{p}(n_j)\}$ drawn as dashed lines.
    The dotted vertical lines mark where the upper and lower interval limits reach
    $\gamma$, giving the range of $n^{*}$ attributable to Monte Carlo error: $[33, 52]$ for Expert~1 and $[53, 64]$ for Expert 2. Because each estimate is based on $500$ simulated tests, the half-width of the interval does not
    exceed approximately 0.04\%, so the conclusions about $n^{*}$ are not driven by simulation noise.
    Near the target, however, the curves are flat, so even this small error moves $n^*$ by several units. The two ranges do not overlap, so the difference between the experts is not an artefact of the simulation.

    \begin{figure}[!t]
    \centering
    \includegraphics[width=0.49\textwidth]{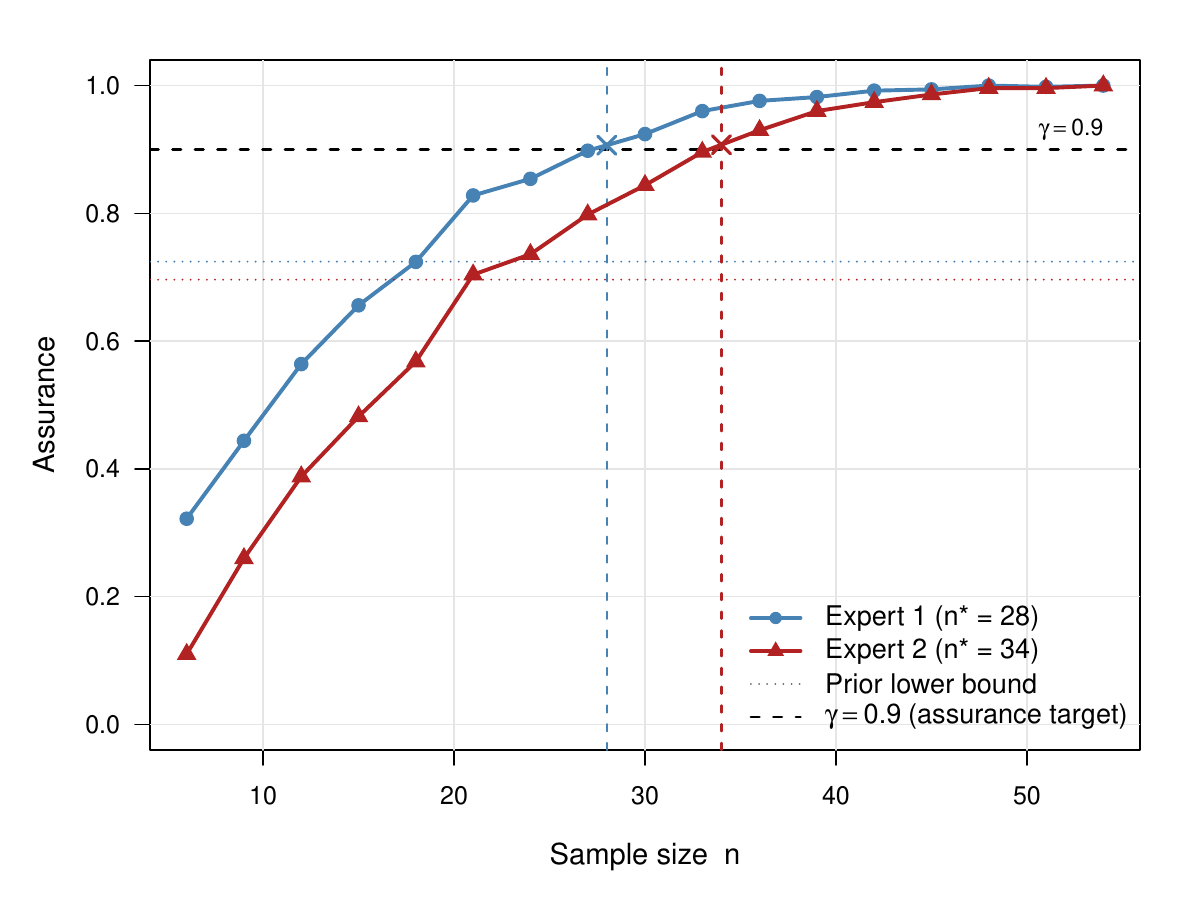}
    \caption{Assurance curves obtained with the empirical approach under the priors of the two experts for a target reliable life of 50 hrs.}
    \label{fig:assurance_E_curves}
    \end{figure}

    For comparison, Figure~\ref{fig:assurance_E_curves} shows the assurance curves obtained with the empirical approach (Figure~\ref{fig:flowchart}). At each $n_j$, $R = 500$ subsamples of $n_j$ units are drawn without replacement from the 90 historical units, with $n_j/3$ units from each stress level. 
    The test data are therefore the observed units themselves, not new data simulated from the Weibull--Eyring model. The resulting assurance is the proportion of $n_j$-unit subsets of the historical data that would pass, and so indicates how many of the observed units would have sufficed. Because the observed sample is treated as the truth, this approach ignores uncertainty in the model parameters and cannot be evaluated beyond the available sample size ($n \le 90$). The empirical curve can therefore approach $1$, and it is closer to a subsampling estimate of power than to Bayesian assurance in the sense of \citet{OHagan2001}.
    
    The choice of methodology has practical consequences for test planning. For Expert~1 at $\tau_{0.90}^{*} = 50$ hrs, the empirical approach gives $n^{*} = 28$, whereas Bayesian assurance gives $n^{*} = 36$; for Expert $2$, the values are $n^* = 34$ and $n^* = 60$. The empirical approach therefore recommends about $22\%$ and $43\%$ of units, respectively. Because it ignores parameter uncertainty, it underestimates the sample size required. A test planned with $n = 28$ would have a Bayesian assurance of only about $0.83$, compared with the target
    $\gamma = 0.90$. The probability that the demonstration fails to confirm the required reliable lifetime therefore rises from $0.10$ to about $0.17$, nearly doubling the risk of an unsuccessful test. Bayesian assurance thus provides a more conservative and more realistic basis for choosing the sample size.
    
    In summary, an assurance of $0.90$ requires $n = 36$ units under the prior of Expert 1 ($n = 60$ under that of Expert $2$). For Expert $1$, $36$ units would be placed on test, $12$ at each of $60$, $72$ and $84\,^{\circ}$C, with $6$ from each location at each temperature, and each unit would be tested until failure or until $t_C = 200$hrs. Given the recorded failure and censoring times, the test is passed if $\Pr\{\tau_{0.90}(60\,^{\circ}\mathrm{C}) \ge 50 \text{ hrs} ~| ~\mathcal{D} \} \ge 0.95$, and failed otherwise. 

    \section{Conclusion}
	\label{sec_7}
    This work develops a Bayesian assurance framework for determining the required sample size in accelerated reliability demonstration testing. The sample size is chosen to achieve a specified probability of successfully demonstrating the reliability requirement, while accounting for uncertainty in the underlying reliability model. A Bayesian Weibull lifetime model is developed within an accelerated life testing framework, with temperature incorporated as the accelerating stress through an Eyring relationship. This enables reliability information obtained at elevated temperatures to be extrapolated to the use condition. Prior distributions for all model parameters are obtained through an elicitation procedure in which expert assessments of observable lifetime quantities are converted into the corresponding hyperparameters.
     
    The framework provides a basis for evaluating alternative sample sizes before testing and identifying a design that achieves the desired level of assurance. Unlike conventional approaches based on a predetermined sample size, the proposed methodology incorporates uncertainty arising from
    the lifetime model, the location effects, and extrapolation from accelerated test conditions into the RDT design.
    
    An additional contribution is the incorporation of a hierarchical location effect within the reliability model. This enables systematic differences between production locations or batches to be represented explicitly, rather than assuming that all observations originate from a single homogeneous population. 

    Bayesian inference is used to estimate the model parameters, with the resulting posterior distributions incorporated into the assurance calculations. For the SCBA data, the test data are informative about all model parameters, and the posteriors obtained under the two experts' priors nearly coincide. The required sample sizes nevertheless differ, because each simulated test is small and is analysed under the expert's prior, which still has a strong influence on the result. The location effect is clearly identified with the full dataset but not with smaller subsamples, so that location heterogeneity adds to the uncertainty of smaller test plans. For an assurance target of $0.9$, the required sample size is sensitive to both the reliability requirement and the elicited prior, with a more diffuse prior leading to a substantially larger test. Planning from the historical data alone underestimates the sample size and makes it more likely that the test fails to demonstrate the requirement. In practice, the sample size for a demonstration test should therefore be chosen by an approach that accounts for uncertainty about the product’s reliability, such as the one proposed here, and its sensitivity to the elicited prior should be checked before testing begins.
    
     The proposed framework is developed for constant-stress tests with Type-I censoring, a single accelerating stress and a Weibull shape parameter that does not vary with stress. Natural extensions include other lifetime distributions and stress--life relationships, and the separate modelling of competing failure modes.    
    The extension of the framework to step-stress accelerated life tests, together with the incorporation of a utility function that accounts explicitly for the costs and benefits associated with alternative testing outcomes and sample sizes, is the subject of ongoing work. 
    
    \bibliographystyle{apalike}
	\bibliography{references}{}
    
    \appendix
    
    \section{Convergence Diagnostics}
    
    \label{app:convergenceA}
    \setcounter{figure}{0}\renewcommand{\thefigure}{A\arabic{figure}}
    \setcounter{table}{0}\renewcommand{\thetable}{A\arabic{table}}
    
    Convergence is assessed under the Expert 1 prior for the full dataset ($n = 90$) and for random subsamples with $2$ and $10$ units per stress level ($n = 6$ and $n = 30$), drawn using \texttt{seed = 123}. Each fit uses three parallel chains with initial values generated randomly from the prior, a burn-in of $1000$ iterations followed by $2000$ iterations per chain, giving $6000$ posterior samples per parameter. The location effects are constrained to sum to zero, so that $\epsilon_2 = -\epsilon_1$. Convergence is assessed using the Gelman--Rubin statistic $\hat{R}$ and its $97.5\%$ upper confidence bound, computed with the \texttt{coda} package using \texttt{autoburnin = FALSE}, the effective sample size (ESS), and trace, density and autocorrelation plots. Results are reported for a single seed at each sample size; the diagnostics are equally satisfactory for subsamples drawn with other seeds and under the Expert 2 prior.
     
     \begin{figure}[tb]
        \centering
        \includegraphics[width=0.70\textwidth]{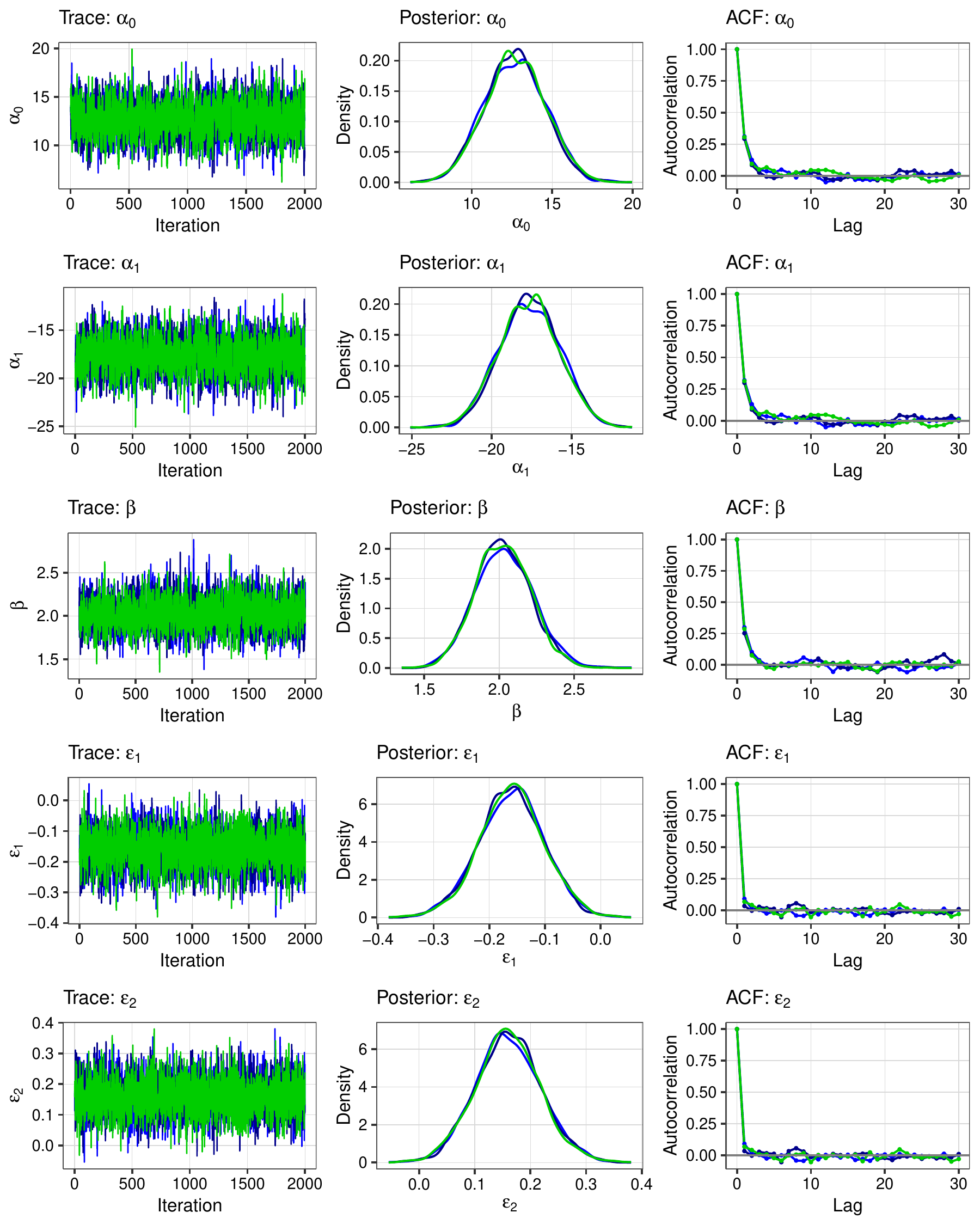}
        \caption{Trace (left), posterior density (centre) and autocorrelation (right) plots for the full dataset ($n = 90$) under the Expert 1 prior. Colours denote the three chains.}
        \label{fig:diagnostics_grid}
    \end{figure} 
    
    Figure \ref{fig:diagnostics_grid} shows these plots for the full dataset. For all five parameters, the three chains fluctuate around a stable level over the iterations, with no trend or drift, and the densities estimated from the individual chains are unimodal and nearly coincide. The autocorrelation of all parameters decays to near zero within five lags. These plots indicate that the burn-in is sufficient, that the three chains have converged to the same distribution and that they mix well.
    
    Table \ref{tab:convergence_by_n} reports the posterior summaries and convergence diagnostics. Across all parameters and sample sizes, $\hat{R} \in [1.000, 1.002]$, with upper confidence bounds in $[1.000, 1.003]$, below the recommended threshold of $1.01$. The smallest ESS in the table is $2737$, well above the commonly recommended minimum of $400$; ESS values slightly above the $6000$ posterior samples reflect mild negative autocorrelation. As $n$ increases, the posterior standard deviation of every parameter decreases, reflecting the increasing information provided by the data.

     \begin{table}[H]
     \centering
        \caption{Posterior summaries and convergence diagnostics under the Expert 1 prior for $n \in \{6, 30, 90\}$. }
        %\label{tab:convergence_by_n}
        \small
        % \resizebox{\textwidth}{!}{%
        \begin{tabular}{rlrrrrrrrrr}
        \hline
        $n$ & Parameter & Mean & SD & 2.5\% & 50\% & 97.5\% & $\hat{R}$ & $\hat{R}_{\text{upper}}$ & ESS  \\
        \hline
        6 & $\alpha_0$ & 15.14 & 3.05 & 9.05 & 15.21 & 20.90 & 1.000 & 1.000 & 3225 \\
         & $\alpha_1$ & $-19.95$ & 2.97 & $-25.56$ & $-20.04$ & $-14.01$ & 1.000 & 1.001 & 3268  \\
         & $\beta$ & 1.62 & 0.40 & 0.94 & 1.59 & 2.48 & 1.001 & 1.002 & 3004 \\
         & $\epsilon_1$ & $-0.23$ & 0.26 & $-0.73$ & $-0.23$ & 0.30 & 1.001 & 1.003 & 2737  \\
         & $\epsilon_2$ & 0.23 & 0.26 & $-0.30$ & 0.23 & 0.73 & 1.001 & 1.003 & 2737  \\
        \hline
        30 & $\alpha_0$ & 14.66 & 2.49 & 9.76 & 14.66 & 19.52 & 1.001 & 1.002 & 3724  \\
         & $\alpha_1$ & $-19.56$ & 2.49 & $-24.39$ & $-19.56$ & $-14.67$ & 1.001 & 1.002 & 3691  \\
         & $\beta$ & 1.61 & 0.24 & 1.17 & 1.60 & 2.12 & 1.002 & 1.003 & 2931  \\
         & $\epsilon_1$ & $-0.21$ & 0.13 & $-0.46$ & $-0.21$ & 0.05 & 1.001 & 1.002 & 6154  \\
         & $\epsilon_2$ & 0.21 & 0.13 & $-0.05$ & 0.21 & 0.46 & 1.001 & 1.002 & 6154  \\
        \hline
        90 & $\alpha_0$ & 12.73 & 1.88 & 9.10 & 12.72 & 16.36 & 1.000 & 1.000 & 3143  \\
         & $\alpha_1$ & $-17.69$ & 1.89 & $-21.33$ & $-17.67$ & $-14.03$ & 1.000 & 1.000 & 3086 \\
         & $\beta$ & 2.02 & 0.18 & 1.68 & 2.02 & 2.39 & 1.001 & 1.001 & 3275  \\
         & $\epsilon_1$ & $-0.16$ & 0.06 & $-0.28$ & $-0.16$ & $-0.04$ & 1.001 & 1.001 & 5726  \\
         & $\epsilon_2$ & 0.16 & 0.06 & 0.04 & 0.16 & 0.28 & 1.001 & 1.001 & 5726 \\
        \hline
        \end{tabular}
        % }
        \label{tab:convergence_by_n}
    \end{table}
     
    The larger posterior uncertainty of the location effects at $n = 6$ reflects the limited information about location heterogeneity when only two units are tested per stress level, but even in this case the diagnostics give no indication of non-convergence. Together with the similar results obtained for other subsamples and under the Expert 2 prior, this indicates that the posterior draws are reliable both for the simulation analysis in Section \ref{subsec:converg_diagno} and for the repeated fits underlying the assurance curve.

    \section{Expert Elicitation Questions} 
    \label{sec:elicitation-protocol}    
    \paragraph{Opening statement:}
    ``I am going to ask for your best judgements about how long these units last. Please give your own view, based on your experience, rather than what you expect the test to show.''
    % -----------------------------
    \subsubsection*{Stage 1: Failure time spread (Weibull shape, $\beta$)}
    \textbf{\textit{Facilitator note:}} For this elicitation, Use $p_1 = \frac{1}{3}$ and $p_2 = \frac{2}{3}$. Stage 1 answers will get $r_1$ from which $\beta$ can be calculated. Using the most frequently used stress condition; e.g. temperature, denoted $t_{\text{ref}}$ :
    \begin{enumerate}[label=\textbf{Q\arabic*.}, leftmargin=*]
        \item Imagine 100 new units all operating continuously at $t_{\text{ref}}$\, until they fail. Approximately how many hours would you expect a typical unit to operate before it fails? In other words, after how many hours would you expect about $50$ out of the $100$ units to have failed? \hfill (median lifetime, $t_{0.5}$)
        \item Thinking about those same $100$ units, by approximately how many hours would you expect the first $33$ units to have failed? ($t_{1/3}$)
        \item By approximately how many hours would you expect 67 units  to have failed? ($t_{2/3}$)
        \item Do failures mostly come from wear-out, or do you also see early or random failures? 
    \end{enumerate}
    
    \textbf{Note}: The smaller the ratio \(r_1\), the greater the separation between the earlier and later failures, indicating a wider spread of lifetimes. 
    % ---------------------------------------------------------------------
    \subsubsection*{Stage 2: Effect of temperature (stress) on lifetime (Eyring parameters, $\alpha_0$ and $\alpha_1$)}
    
         \textbf{\textit{Facilitator note:}} Stage 2 answers will get $r_2$. Consider three temperature values; the normal use, intermediate and higher temperatures. We need to determine the medians ($M_{s_i} /M_{s_j}$) 
    \begin{enumerate}[label=\textbf{Q\arabic*.}, leftmargin=*, resume] 
         \item At the normal operating temperature, $t_{\text{ref}}$, approximately how many hours would you expect a typical unit to last before it fails? (\(M_{t_{\mathrm{ref}}}\)) \hfill (median at use case stress)  
         \item If the temperature were increased to $t_{\text{mid}}$, approximately how many hours would you expect a typical unit to last before it fails? (\(M_{t_{\mathrm{mid}}}\)) \hfill (median at intermediate stress)
         \item If the temperature were increased further to $t_{\text{high}}$, approximately how many hours would you expect a typical unit to last before it fails? (\(M_{t_{\mathrm{high}}}\))\hfill (median at highest stress)
         \item Approximately how many hours would you expect to pass before 63 out of the 100 units have failed? \textbf{\textit {Facilitator Note:}} This questions is specifically for the reference batch, for which \(\epsilon_l=0\).
        \item Roughly how much shorter would you expect the typical lifetime to be at the highest test temperature? For example, would you expect the equipment to last half as long, one-quarter as long, or even less than at the use condition? (Reasonableness check on the medians)

    \end{enumerate}
    % ---------------------------------------------------------------------
    \subsubsection*{Stage 3: Uncertainty in the median lifetimes (covariance, $\boldsymbol{\Sigma}_{\alpha}$)}    
      Questions below are asked separately at each of the different stress conditions in this paper it is three temperatures $t_{\text{ref}}$, $t_{\text{mid}}$ and $t_{\text{high}}$, where $m_t$ denotes the median lifetime the expert gave for temperature in Stage~2.
    
    \begin{enumerate}[label=\textbf{Q\arabic*.}, leftmargin=*, resume]
        \item You said a typical unit lasts about $m_t$ hours at a given temperature.  Please give me a \emph{lower} value such that you think there is only a 1-in-4 chance that the true typical lifetime is below it.($M_{t,0.25}$) \hfill (lower quartile)
        \item Now please give me a \emph{higher} value such that you think there is only a 1-in-4 chance that the true typical lifetime is above it. $M_{t,0.75}$ \hfill (upper quartile)
        \textbf{\textit{Facilitator note:}} confirm that the expert still regards $m_t$ as the value for which the true typical lifetime is equally likely to be higher or lower.
    
        \item ``If your estimate at $T_{\text{high}}$ turned out to be too optimistic, would your estimate at $T_{\text{use}}$ probably also be too optimistic? 
    \end{enumerate}
    % ---------------------------------------------------------------------
    \subsubsection*{Stage 4: Differences between test locations (location random effect)}
    \begin{enumerate}[label=\textbf{Q\arabic*.}, leftmargin=*, resume]
        \item ``Think of the best and the worst [site / batch / test facility]. Under the same conditions, how much longer would a typical unit from the best one last than a typical unit from the worst? For example, twice as long, or about 10\% longer?''
        \item ``Would that difference be the same at every temperature?''
    \end{enumerate} 

\end{document}

%% file: flowchart_tikz.tex
% Flowchart of the assurance procedure (two routes for the test data)
% Used in the main file with \input{flowchart_tikz} inside a figure environment.
% \usepackage{tikz} and \usetikzlibrary{shapes.geometric, arrows.meta, positioning, calc}

\begin{tikzpicture}[
    node distance = 6mm and 5mm,
    box/.style  = {rectangle, draw, thick, rounded corners=2pt, align=center,
                   text width=2.3cm, minimum height=13mm, font=\footnotesize},
    gen/.style  = {box, text width=2.9cm},
    data/.style = {box, fill=black!6},
    arr/.style  = {-{Latex[length=1.8mm]}, thick},
]

% --- inputs ---
\node[data] (expert) {Expert\\ judgements};
\node[data, below=of expert] (hist) {Historical ALT data $\tilde{\mathcal{D}}$};

% --- prior and design posterior ---
\node[box, right=of expert] (prior) {Elicited prior\\ $p(\boldsymbol{\theta})$};
\node[box, right=of hist] (design) {Design posterior\\ $p(\boldsymbol{\theta} \mid \tilde{\mathcal{D}})$};

% --- test data for replication r: (a) model-based, (b) resampling ---
\node[gen, right=9mm of design] (genG)
      {\textbf{Proposed} \\Draw $\boldsymbol{\theta}^{(w)} \sim p(\boldsymbol{\theta} \mid \tilde{\mathcal{D}})$;
       simulate $\mathcal{D}^{(w)}$ of size $n_j$};
\node[gen, below=of genG] (genE)
      {\textbf{Empirical}\\ Subsample $\mathcal{D}^{(w)}$ of size $n_j$ from $\tilde{\mathcal{D}}$};

% --- fit and decision ---
\coordinate (mid) at ($(genG.east)!0.5!(genE.east)$);
\node[box, right=7mm of mid, anchor=west] (fit) {Fit model to $\mathcal{D}^{(w)}$ under $p(\boldsymbol{\theta})$};
\node[box, right=of fit] (pass) {$x^{(w)} = 1$ if the test passes (Eq.~\ref{eq:decision_criterion})};

% --- outputs ---
\node[box, below=11mm of pass |- genE.south, anchor=north] (assur) {Assurance $\hat{p}(n_j)$ with 95\% CI};
\node[data, left=of assur] (nstar) {$n^*$ and its range};

% --- loop frame ---
\draw[dashed, rounded corners=4pt]
      ($(genG.north west)+(-0.25,0.3)$) rectangle ($(pass.east |- genE.south)+(0.25,-0.65)$);
\node[anchor=south west, font=\footnotesize\itshape]
      at ($(genE.south west)+(-0.2,-0.6)$) {repeat $W$ times for each $n_j$};

% --- arrows ---
\draw[arr] (expert) -- (prior);
\draw[arr] (hist) -- (design);
\draw[arr] (prior) -- (design);
\draw[arr] (design) -- (genG);
\draw[arr] (hist.south) |- (genE.west);
\draw[arr] (genG.east) -- ++(0.3,0) |- (fit.west);
\draw[arr] (genE.east) -- ++(0.3,0) |- (fit.west);
\draw[arr] (prior.north) -- ++(0,0.35) -| (fit.north);
\draw[arr] (fit) -- (pass);
\draw[arr] (pass) -- (pass |- assur.north);
\draw[arr] (assur) -- (nstar);

\end{tikzpicture}